\documentclass[twocolumn]{autart}    

\usepackage{graphicx}          

\usepackage{tikz}
\usepackage{pgfplots}
\pgfplotsset{compat=newest}
\usepgfplotslibrary{external} 
\usetikzlibrary{plotmarks}
\usetikzlibrary{arrows.meta}
\usepgfplotslibrary{patchplots}
\usepackage{grffile}
\usepackage{amsmath}
\usepackage{mathtools} 
\usepackage{amssymb}
\usepackage{dsfont}
\usepackage{algorithm}
\usepackage{algpseudocode}
\usepackage{multirow}
\usepackage{xcolor}
\usepackage{array}
\usepackage{csquotes}

\makeatletter
\newcommand\norm[1]{\left\lVert#1\right\rVert}

\usepackage{xpatch}
\makeatletter 
\xpatchcmd{\runningauthor@fmt}{\global\edef}{\protected@xdef}{}{}
\xpatchcmd{\runningauthor@fmt}{\global\edef}{\protected@xdef}{}{}
\xpatchcmd{\author@fmt}{\edef}{\protected@edef}{}{}
\def\@xnamedef#1{\expandafter\protected@xdef\csname #1\endcsname}
\def\ead@au#1{\protected@edef\@ead@au{#1}}
\def\add@xtok#1#2{\begingroup
  \protected@xdef\@act{\global\noexpand#1{\the#1#2}}\@act
\endgroup}
\def\no@harm{}

\begin{document}

\begin{frontmatter}

\title{Robust Correlated Equilibrium: Definition and Computation\thanksref{footnoteinfo}} 

\thanks[footnoteinfo]{This paper was not presented at any IFAC 
meeting. Corresponding author: Rahul Misra}

\author[AAU]{Rahul Misra}\ead{rmi@es.aau.dk},    
\author[AAU]{Rafał Wisniewski}\ead{raf@es.aau.dk},               
\author[AAU]{Carsten Skovmose Kallesøe}\ead{csk@es.aau.dk},  
\author[UCL]{Manuela L. Bujorianu}\ead{l.bujorianu@ucl.ac.uk}  

\address[AAU]{Automation and Control section, Department of Electronic Systems, Aalborg
University, 9220 Aalborg East, Denmark}  
\address[UCL]{Department of Computer Science, University
College London, United Kingdom}             


\begin{abstract}                          
    We study N-player finite games with costs perturbed due to time-varying disturbances in the underlying system and to that end, we propose the concept of Robust Correlated Equilibrium that generalizes the definition of Correlated Equilibrium. Conditions under which the Robust Correlated Equilibrium exists are specified, and a decentralized algorithm for learning strategies that are optimal in the sense of Robust Correlated Equilibrium is proposed. The primary contribution of the paper is the convergence analysis of the algorithm and to that end, we propose a modification of the celebrated Blackwell's Approachability theorem to games with costs that are not just time-average, as in the original Blackwell's Approachability Theorem, but also include the time-average of previous algorithm iterates. The designed algorithm is applied to a practical water distribution network with pumps being the controllers and their costs being perturbed by uncertain consumption due to the consumers. Simulation results show that each controller achieves no regret, and empirical distributions converge to the Robust Correlated Equilibrium. 
\end{abstract}

\end{frontmatter}

\section{Introduction}
Over the last few decades, the scope of control applications has shifted from optimizing decisions for small-scale systems to large-scale systems with multiple decision-makers \cite{annaswamy2023control}. To ensure scalable control design for such large-scale systems, decentralized control schemes are required, and game theory is used to that end \cite{mageirou1977decentralized,marden2018game}. Game theory is a mathematical framework used to study the interactions between multiple decision-makers (also known as players), where a decision made by one player affects not only his/her costs but also the costs of other players. Possible outcomes of the game are characterized by a game-theoretic equilibrium where no player benefits by unilateral deviation of his/her decision. Since the inception of equilibrium concepts such as Nash Equilibrium ($\mathcal{NE}$) \cite{nash1951}, it has been argued whether the strategies used by individual rational players will eventually converge to an $\mathcal{NE}$. One challenge with computing a general $\mathcal{NE}$ is that no algorithm can be designed for computing it in \emph{polynomial-time} \cite{shoham2009multiagent} i.e. the problem scales badly when the number of players or the number of control actions is increased which is a hallmark of large-scale systems. Another fundamental challenge in game theory is \emph{how do players reach an equilibrium?} Fictitious play was introduced as the first algorithm \cite{brown1951iterative} to answer that question. Unfortunately, fictitious play can only be used for zero-sum, cooperative games or games with players having at most $2$ actions \cite{benaim1999mixed,fudenberg1998theory}, and there do not exist any simple decentralized algorithms that ensure convergence to a general $\mathcal{NE}$ \cite{hart2003uncoupled}. Furthermore, the concept of $\mathcal{NE}$ is incompatible with the Bayesian view of probability, and a player may not choose to follow the $\mathcal{NE}$ strategy after observing the opponent's actions since $\mathcal{NE}$ assumes the worst-case move by opponents \cite{aumann1987correlated}. Correlated Equilibrium ($\mathcal{CE}$) which is a convex combination of $\mathcal{NE}$ points is considered to be a more natural alternative to $\mathcal{NE}$ as it assigns subjective probabilities to all possible outcomes of the game \cite{aumann1974subjectivity,aumann1987correlated} (unlike worst-case outcome in $\mathcal{NE}$) as well as is easier to compute as it is a convex polytope \cite{papadimitriou2008computing}. The celebrated Regret matching Algorithms allow decentralized learning of a $\mathcal{CE}$ by a group of agents \cite{hart2000simple,hart2001reinforcement,hart2013simple}. In this paper, we aim to extend the concept of $\mathcal{CE}$ to games with costs perturbed by disturbances or the so-called time-varying games that have gained a lot of research interest recently \cite{zhang2022no,wang2023convergence,feng2023last}. To the best of our knowledge, there is a lack of literature on time-varying general games. A related setting is the \emph{contextual games} introduced in \cite{sessa2020contextual}, where the disturbance can be thought of as a context. However, unlike \cite{sessa2020contextual}, in this work, the players do not observe the disturbance, and we focus on the stricter notion of Robust $\mathcal{CE}$ (instead of coarse-correlated equilibrium in \cite{sessa2020contextual}). The motivation behind this work is from practical decentralized optimal control applications where costs could be perturbed by disturbances or unknown dynamics in large-scale cyber-physical systems such as traffic routing, smart grids, pandemics, or energy infrastructure such as water distribution networks \cite{annaswamy2023control,karaca2020no,misra2023decentralized}. 

\textbf{Summary of contributions}: 1. We propose a modification of $\mathcal{CE}$ for time-varying games with finite control actions and finite disturbances in the most general uncoupled setting and prove its existence, 2. We propose an algorithm and its convergence analysis for computing the solution, 3. The convergence analysis of the proposed algorithm required a modification of Blackwell's Approachability theorem for the time average of costs as well as the time average of previous iterates of the proposed algorithm, 4. Lastly, we apply our algorithm for decentralized control of a water distribution network. \newline
\textbf{Paper Outline}: We introduce the notation and thereafter introduce the problem setup and our solution concept in Section 3. The existence of the solution is proven in Section 4 followed by an Algorithm with convergence analysis in Section 5. Simulation studies are presented in Section 6 followed by conclusions in Section 7.     

\section{Notation}
\begin{itemize}
    \item For a finite set $A$, $\mid A \mid$ is the cardinality of $A$.
    \item $\mathds{P}[\mu]$ represents the probability of event $\mu$ occurring.
    \item $\mathds{E}_{\mu}$ represents the expectation operator corresponding to the probability measure $\mu$ in the subscript.
    \item $\Delta(A)$ represents a probability simplex in $\mathds{R}^{\mid A \mid}$ where $A$ is a finite set. Formally
    \begin{equation*}
        \Delta(A) := \left\{x \in \mathds{R}_+^{\mid A \mid}: \sum\limits_{a \in A} x(a) = 1 \right\}.
    \end{equation*}
    \item $[a]_+$ represents nonnegative part of the real number $a$ i.e. $[a]_+ = \max\{a,0\}$ and for a vector of reals $a = (a_1, \cdots, a_m)$, represents $[a]_+ = ([a_1]_+, \cdots, [a_m]_+)$ and similarly for a matrix $A$ of reals, $[A]_+$ represents matrix consisting of entries $[a_{ij}]_+$.  
    \item  $\norm{a}_{p}$ denotes the $p$-norm of the vector $a \in \mathds R^n$. If no subscript is present, it represents the $2$-norm.
    \item For two vectors $a \in \mathbb{R}^n$ and $b \in \mathbb{R}^n$, $a\cdot b$ represents the standard scalar product between them. 
    \item Given $u = (u^i, u^{-i})$, the notation $\sum\limits_{u \in U: u^i = a}(\cdot)$ is a conditional sum where $u^i$ is fixed as $a$.
\end{itemize}
\section{Problem setup and formulation}
The considered problem setup in this work is similar to \cite{hart2013simple}, but differs from \cite{hart2013simple} in the following ways: 1. The costs will be perturbed by an unknown (possibly adversarial) disturbance, and therefore, the costs and all associated variables will be time-dependent, 2. Each player can only observe the costs incurred by them (they do not have a cost function for calculating their costs but simply incur a cost at each time instant from the environment) and cannot observe the control actions taken by the other players (each player won't be even aware that they are playing a game and that there are other players i.e. \emph{Unkown Game} setting). We begin by defining a static game $\Gamma$.         
\subsection{Static game preliminaries} \label{subsection:static game}
Consider a $N$-player game $\Gamma$ in strategic (normal) form. The game $\Gamma$ is defined as a tuple $\Gamma = ( N,U,{C})$, where 
\begin{equation*}\label{eq:ActionSpace}
    U := \prod\limits_{i \in \{1, \cdots, N\}} U^i,
\end{equation*} 
is the space of finite joint control actions for all players with $U^i$ being the finite control action space for the player $i$ and ${C} = \{c^1, \cdots, c^N\}$ are the costs incurred by the players due to the joint control actions $U$ taken by all the players. Specifically for any player $i \in \{1, \cdots, N\}$, $c^i:U \mapsto \mathds{R}$. Define $U^{-i}$ as
\begin{equation*}\label{eq:ActionSpace_-i}
    U^{-i} := \prod\limits_{j \in \{1, \cdots, N\}, j \neq i} U^j,
\end{equation*}
which represents the joint control action space of all players except player $i$. In this paper, $u \in U$ will represent a joint control action i.e. $u = (u^i, u^{-i})$. Since an equilibrium may not exist in the space of pure control actions but always exists in the space of randomized (or mixed) control actions \cite{nash1951}, we define a randomized control policy. A randomized policy for a player $i \in \{1, \cdots, N\}$ is a probability distribution $x^i \in \Delta(U^i)$ over player $i$'s pure control actions $U^{i}$. A randomized joint distribution is a probability distribution $x \in \Delta(U)$ over all the player's joint pure control actions $U$. Given the joint distribution $x$, we can obtain the marginal distributions $x^i \in \Delta(U^i)$ as $x^i(u^i) = \sum_{{u}^{-i} \in U^{-i}} x (u^i, u^{-i})$, $\forall u^i \in U^{i}$ and $x^{-i} \in \Delta(U^{-i})$ as $x^{-i}(u^{-i}) = \sum_{{u}^{i} \in U^{i}} x (u^i, u^{-i}) \quad \forall u^{-i} \in U^{-i}$. The domain of cost functions can be extended multilinearly from joint control action space $U$ to the joint probability distribution $\Delta(U)$ as follows 
\begin{equation*}
    c(x) := \mathds{E}_{x} [c] = \sum\limits_{u \in U} c(u)x_u.
\end{equation*}
Similarly, we can write the costs associated with the marginal distributions $x^i$ or $x^{-i}$.
\begin{defn}\label{defn:Correlated_Equilibrium}
    A probability distribution $\psi \in \Delta(U)$ is called a $\mathcal{CE}$ of $\Gamma$ if it satisfies the following condition for each player $i \in \{1, \cdots, N\}$ and every $a, b \in U^i$,
    \begin{equation} \label{eq:Correlated_Equilibrium}
        \sum\limits_{u \in U: u^i = a} \psi(u)[c^i(a,u^{-i}) - c^i(b,u^{-i})] \leq 0,
    \end{equation}
\end{defn}
 The $\mathcal{CE}$ condition \eqref{eq:Correlated_Equilibrium} should be interpreted as follows: Suppose a mediator ($\psi$) for the considered game $\Gamma$ recommends action $a$ to player $i \in \{1, \cdots, N\}$ i.e. $u^i = a$. Then after observing the recommendation, it is beneficial for player $i$ to follow the recommendation $u^i = a$ i.e. the costs incurred by player $i$ increase if player $i$ deviates from the recommendation $u^i = a$. Specifically, for a two-player game, for every $a, b \in U^1$ and $\beta \in U^2$,
   \begin{equation*} 
        \sum\limits_{\beta \in U^2} \psi(a,\beta)[c^1(a,\beta) - c^1(b,\beta)] \leq 0, 
    \end{equation*}
    and for every $e, f \in U^2$ and $\alpha \in U^1$,
       \begin{equation*} 
        \sum\limits_{\alpha \in U^1} \psi(\alpha,e)[c^2(\alpha,e) - c^2(\alpha,f)] \leq 0. 
    \end{equation*}
    Thus the distribution $\psi$ is conditioned on the recommendation observed by the player $i$ for any player $i \in \{1, \cdots, N\}$. The set of $\mathcal{C}\mathcal{E}$ with cost functions $c: U \to \mathds R^N$ is defined as follows
    \begin{multline}\label{eq:CE_set}
        \mathcal{C}\mathcal{E}(c) = \bigg\{ \psi \in \Delta(U): \sum\limits_{u \in U: u^i = a} \psi(u)c^i(a,u^{-i}) \leq \\  \sum\limits_{u \in U: u^i = a} \psi(u)c^i(b,u^{-i}), \quad \forall i, a, b \bigg\} 
    \end{multline}
By Definition \ref{defn:Correlated_Equilibrium}, the correlated equilibrium is robust to any possible unilateral deviation from any player $i \in \{1, \cdots, N\}$ (in expectation with respect to the distribution $\psi(u)$). However, what happens if the costs are perturbed due to exogenous disturbances? Is the equilibrium condition \eqref{eq:Correlated_Equilibrium} still robust to perturbations in the cost function itself due to unknown disturbances? This is the primary focus of this paper, and in the following subsection, we define this setting more precisely.   

\subsection{Static Game with finite Perturbed costs}
Suppose that the costs incurred by each player in the Static Game $\Gamma$ are perturbed due to disturbances. We assume that the number of possible disturbances $D$ is finite, and this implies that a finite number of perturbed costs exist such that the disturbance picks one of the costs (from a finite set of costs). Such a Static game will be referred to as a Perturbed Static game. Let $D$ be the finite number of costs then formally we define an $N$-player Perturbed Static game as $\Gamma' = ( N, U, {C}_D)$ where $U$ is the joint control space, and $C_D: U \to \mathds R^{N \times D}$ maps a control action to a $N \times D$ matrix. Each $c^i: U \to  \mathds R^D$ is the $N^{th}$ row of the matrix map $C_D$ with entries $c^i_d$ where $d \in \{1, \cdots, {D}\}$ (unlike scalar costs in subsection \ref{subsection:static game}). We now extend the definition of Correlated equilibrium for Perturbed Static games by introducing the concept of Robust Correlated Equilibrium ($\mathcal{RCE}$).     
\begin{defn}
     A joint distribution $\Psi$ is called an $\mathcal{R}\mathcal{C}\mathcal{E}$ for the Perturbed Static Game $\Gamma'$ \textcolor{black}{if and only if} it is an element of the following set
    \begin{multline}\label{eq:RCE_set}
        \mathcal{R}\mathcal{C}\mathcal{E} = \bigg\{ \Psi \in \Delta(U): \sum\limits_{u \in U: u^i = a} \Psi(u)c^i_d(a, u^{-i}) \leq \\ \sum\limits_{u \in U: u^i = a} \Psi(u)c^i_d(b,u^{-i}),\text{ }\forall i \in \{1, \cdots, N\}, \\ \forall a \in U^i, b \in U^i, \text{ and } \forall d \in \{1, \cdots, {D}\} \bigg\}. 
    \end{multline}
\end{defn}
It should be noted that the $\mathcal{RCE}$ condition is a stricter notion compared to the standard Correlated Equilibrium condition \eqref{eq:CE_set} as 
\begin{equation}\label{eq:RCE_short}   
    \mathcal{R}\mathcal{C}\mathcal{E} = \underset{d = 1}{\overset{D}{\bigcap}}\mathcal{C}\mathcal{E}(c_d), 
\end{equation}
where $c_d: U \to \mathds R^N$ is the $D^{th}$ column of matrix map $C_D$. \textcolor{black}{Just as the standard $\mathcal{CE}$ is a convex combination of Nash equilibrium ($\mathcal{NE}$) points \cite{shoham2009multiagent,aumann1974subjectivity,aumann1987correlated}, $\mathcal{RCE}$ is a convex combination of robust $\mathcal{NE}$ points, as \eqref{eq:RCE_set} is a convex set since it is an intersection of convex sets \cite{boyd2004convex}.} More compactly, the $\mathcal{RCE}$ condition can be stated as a probability distribution $\Psi \in \Delta(U)$ such that, for each player $i \in \{1, \cdots, N\}$, with cost $c^i: U \to \mathds R^D$ and for every $a, b \in U^i$, the following condition holds
\begin{multline} \label{eq:Robust_Correlated_Equilibrium}
    \sum\limits_{u \in U: u^i = a} \Psi(u)[ c^i_d(a,u^{-i}) - c^i_d(b,u^{-i})] \leq 0, \\ \forall d \in \{1, \cdots, {D}\}. 
\end{multline}
Furthermore, as $\mathcal{RCE}$ is a closed subset of a bounded set $\Delta(U)$, it is compact. Unlike correlated equilibrium, the existence of $\mathcal{RCE}$ i.e. non-emptiness of \eqref{eq:RCE_set} is non-trivial as shown by the following simple example.
    
\subsection{Motivation for Robustifying $\mathcal{CE}$} \label{subsection:Motivation}

We present a simple resource-sharing problem. Consider two farms that use water from a common irrigation channel. Each farm has an automatic controller that directs water from the irrigation channel to the farm. Assume, for simplicity, that both the farms continuously require water. The controller can either \emph{open} the valve (action $O$) or keep it \emph{closed} (action $C$). Under all conditions, the water in the irrigation channel can supply only one farm at a time. If both farms try to use water from the irrigation channel simultaneously, it will be drained for a long time. The cost incurred by the controllers for opening the valve simultaneously is $8$ units. The cost incurred by the controller for opening the valve while the other controller's valve is closed is $0$ units and the cost incurred for keeping the valve closed while the other farm consumes the water is $5$ units. Lastly, if both the controllers do not consume water they pay $1$ unit each. The water available in the common irrigation supply depends on the weather conditions (specifically, whether there is a drought). Under drought conditions, it becomes more expensive for the farm to not consume water while the other farm consumes it and this is reflected in the costs incurred by the controllers as the cost incurred by a controller for keeping the valve closed, while the other controller opens its valve is increased to $7.5$ units form $5$ units. The following cost matrices summarize the game. 
\begin{table}[ht] \label{tb:CEQ_table}
\centering
    \text{$d =$ Normal}
    \setlength{\extrarowheight}{7pt}
    \begin{tabular}{*{4}{c|}}
      \multicolumn{2}{c}{} & \multicolumn{2}{c}{Control $2$}\\\cline{3-4}
      \multicolumn{1}{c}{} &  & $C$  & $O$ \\\cline{2-4}
      \multirow{2}*{Control $1$}  & $C$ & $(1, 1)$ & $(5,0)$ \\\cline{2-4}
      & $O$ & $(0,5)$ & $(8,8)$ \\\cline{2-4}
    \end{tabular}
 \end{table}
 \begin{table}[ht]
\centering
    \text{$d =$ Drought}
    \setlength{\extrarowheight}{7pt}
    \begin{tabular}{*{4}{c|}}
      \multicolumn{2}{c}{} & \multicolumn{2}{c}{Control $2$}\\\cline{3-4}
      \multicolumn{1}{c}{} &  & $C$  & $O$ \\\cline{2-4}
      \multirow{2}*{Control $1$}  & $C$ & $(1, 1)$ & $(7.5,0)$ \\\cline{2-4}
      & $O$ & $(0,7.5)$ & $(8,8)$ \\\cline{2-4}
    \end{tabular}
 \end{table}
 Let $\psi_{CC}, \psi_{CO}, \psi_{OC}$, and $\psi_{OO}$ represent the probabilities of the recommender suggesting the actions $\{C, C\}$, $\{C, O\}$, $\{O, C\}$, and $\{O, O\}$ to controller $1$ and $2$ respectively. Consider the correlated equilibrium where the public recommendation to both players is as follows,
 \begin{equation}\label{eq:CEQ_example}
     \psi_{CC} = \frac{1}{3}, \quad, \psi_{CO} = \frac{1}{3}, \quad \psi_{OC} = \frac{1}{3}, \quad \psi_{OO} = 0.
 \end{equation}
  Consider the normal conditions i.e. ($d =$ Normal condition), and suppose that the control $1$ is recommended action $C$, than control $1$ can infer that control $2$ must have received the recommendation $C$ or $O$ with equal probability and therefore the expected cost for control $1$ is,
  \begin{align*}
     &\frac{1}{2}(1) + \frac{1}{2}(5) = 3, \text{ if control $1$ follows $C$}, \\
     &\frac{1}{2}(0) + \frac{1}{2}(8) = 4, \text{ if control $1$ switches to action $O$},
  \end{align*}
   and under recommendation $O$, the control $1$ infers that control $2$ must have been recommended to choose $C$ with probability $1$, and therefore cost incurred by control $1$ is $0$ for following the recommendation and $1$ unit for not following the recommendation. Thus, the distribution \eqref{eq:CEQ_example} is a correlated equilibrium. Now consider the drought conditions i.e. ($d = $ Drought), and suppose that the control $1$ is recommended action $C$, than control $1$ can infer that control $2$ must have received the recommendation $C$ or $O$ with equal probability and therefore the expected cost for control $1$ is,
  \begin{align*}
     &\frac{1}{2}(1) + \frac{1}{2}(7.5) = 4.25, \text{ if control $1$ follows $C$}, \\
     &\frac{1}{2}(0) + \frac{1}{2}(8) = 4, \text{ if control $1$ switches to action $O$},
  \end{align*}   
and the costs remain the same if the recommendation to control $1$ was $O$. Clearly, it is better for control $1$ to not follow the recommendation in the case recommendation is $C$ during drought conditions and this demonstrates that the correlated equilibrium distribution \eqref{eq:CEQ_example} is not robust to the disturbance $d$. \textcolor{black}{More abstractly in this example, one can derive the threshold of robustness by solving the following set of linear inequalities. For any action $a, b \in U^1 = \{C, O \}$, with disturbances $d_1, d_2$, we need to simultaneously  satisfy,
    \begin{align*}
        &\psi(a,C)\left[c^1_{d_1}(a,C) - c^1_{d_1}(b,C)\right] \\
        &\quad \quad \quad \quad + \psi(a,O)\left[c^1_{d_1}(a,O) - c^1_{d_1}(b,O)\right] \leq 0, \\
        &\psi(a,C)[c^1_{d_2}(a,C) - c^1_{d_2}(b,C)] \\
        &\quad \quad \quad \quad  + \psi(a,O)\left[c^1_{d_2}(a,O) - c^1_{d_2}(b,O)\right] \leq 0.
    \end{align*}
    and for every $e, f \in U^2 = \{C,O\}$, with disturbances $d_1, d_2$, we also need to satisfy,
    \begin{align*}
        &\psi(C,e)\left[c^2_{d_1}(C,e) - c^2_{d_1}(C,f)\right] \\
        &\quad \quad \quad \quad + \psi(O,e)\left[c^2_{d_1}(O,e) - c^2_{d_1}(O,f)\right] \leq 0, \\
        &\psi(C,e)\left[c^2_{d_2}(C,e) - c^2_{d_2}(C,f)\right] \\
        &\quad \quad \quad \quad  + \psi(O,e)\left[c^2_{d_2}(O,e) - c^2_{d_2}(O,f)\right] \leq 0.
    \end{align*}
    }

\section{Existence of $\mathcal{RCE}$}
\textcolor{black}{This section focuses on the question of the existence of $\mathcal{RCE}$ for a game. We now state Helly's Theorem from discrete convex geometry. Helly's Theorem provides conditions for the non-emptiness of the intersection of a family of convex sets. In our case, the family of convex sets are $\mathcal{CE}$ parameterized by the perturbed costs $c_d$ \eqref{eq:CE_set}.  
\begin{thm}[Helly's Theorem \cite{matousek2013lectures}]\label{th:Helly}
    Let $\mathcal{A}$ be an arbitrary infinite family of compact convex sets in $\mathbb{R}^n$ such that any $n + 1$  of the sets have a non-empty intersection. Then all the sets of $\mathcal{A}$ have a nonempty intersection, i.e., $\underset{i = 1}{\overset{\infty}{\bigcap}}\mathcal{A}_i \neq \emptyset$.
\end{thm}
The following Proposition states the existence of $\mathcal{RCE}$. 
\begin{prop}\label{th:existence}
    A finite perturbed game $\Gamma'$ has an $\mathcal{RCE}$ as per \eqref{eq:RCE_set} if and only if the intersection of every $\mid U \mid$ sets of $\mathcal{CE}$ is non-empty. 
\end{prop} 
\begin{pf}
  Without loss of generality, we suppose that $D \geq \mid U \mid$ (by concatenating the perturbed costs $(c_1, \cdots, c_D)$ with additional  $D-\mid U \mid$ repetitions of the last element $c_D$). Recall that $\mathcal{RCE}$ is the intersection of compact, convex sets of $\mathcal{CE}$ parameterized by the perturbed costs $c_d$ (see \eqref{eq:RCE_short}). 
  Our goal is to prove the following: 
   For all $\mathcal{N} \subset \{1, \cdots, D\}$ if the following condition 
    \begin{equation*}
        \mid \mathcal{N} \mid = \mid U \mid, \quad \underset{{d \in \mathcal{N}}}{\bigcap} \mathcal{CE}(c_d) \neq \emptyset
    \end{equation*}
    is satisfied then,
    \begin{equation*}  
        \mathcal{R}\mathcal{C}\mathcal{E} = \underset{d = 1}{\overset{D}{\bigcap}}\mathcal{C}\mathcal{E}(c_d) \neq \emptyset.
    \end{equation*} 
    As per \eqref{eq:RCE_set}, the set $\mathcal{RCE}$ is defined on the simplex $\Delta(U)$. The simplex $\Delta(U)$ is the positive orthant of dimension $\mid U  \mid - 1$ as $\Psi(u) \geq 0$ for all $u \in U$, and $\sum_{u \in U} \Psi(u) = 1$ constrains the value of the last element of $\Psi(u)$ based on previous $\mid U  \mid - 1$ elements. Consider the family of convex sets $\mathcal{A} := \mathcal{CE}(c_1), \cdots, \mathcal{CE}(c_D)$ in $\mathbb{R}^{\mid U \mid - 1}$. Since the number of disturbances $D$ is finite, the family of convex sets $\mathcal{A}$ will be compact. Suppose that the intersection of every $\mid U \mid$ of these sets is non-empty, then by Theorem \ref{th:Helly}, the entire intersection $\mathcal{R}\mathcal{C}\mathcal{E} = {\bigcap}^D_{d = 1}\mathcal{C}\mathcal{E}(c_d) \neq \emptyset$ \qed
\end{pf}
}
\section{Algorithm for $\mathcal{RCE}$}
We have now established the existence of $\mathcal{RCE}$ and this section aims to construct an algorithm that players can use to learn their optimal strategies in the \emph{Unknown Game} setting (i.e., decentralized computation with players unaware of other players' existence in the environment) such that the joint distribution of all the players converges to a $\mathcal{RCE}$. We shall now introduce the learning problem in the \emph{Unknown Game} setting as a repeated game (in subsection \ref{subsection:Setting}), \textcolor{black}{followed by an introduction to the concepts of \emph{Enforceability} and \emph{Approachability}, culminating with the statement of Blackwell's Approachability Theorem (in subsection \ref{subsection:Algorithm}), and the Regret Matching Algorithm from \cite{hart2000simple}. The Regret Matching Algorithm is a consequence of Blackwell's Approachability Theorem. It is a decentralized algorithm where every player minimizes their own internal regrets based on the history of play. If every player updates their mixed strategy based on Regret Matching, then the empirical joint distribution of play is guaranteed to converge (almost surely) to the set of Correlated Equilibrium (Theorem A in \cite{hart2000simple}). However, in the case of Perturbed Games, we cannot directly apply the standard Regret Matching Algorithm as it does not take into account the perturbed costs $c_d$. Therefore, we introduce the notion of Perturbed Conditional Regrets, and the proposed algorithm is essentially Regret Matching applied on Perturbed Regrets.} 
\subsection{Setting}\label{subsection:Setting}
We begin by formalizing the learning problem as a Repeated Game with Perturbed Costs where the aforementioned Static Game with Perturbed costs is repeated at each time instant: $t = 1,2, \cdots$. Each player $i \in \{1, \cdots, N\}$ observes the history of play as 
\begin{equation*}
    h^i_t = [u^i_1, c^i_{d_1}, u^i_2, c^i_{d_2} \cdots, u^i_t, c^i_{d_t}],
\end{equation*}
where $u^i_t$ and $c^{i}_{d_t}$ are defined as the control action taken by player $i$ at time $t$ (indicated by the subscript), and $c^{i}_{d_t}$ is the cost incurred by player $i$ at time $t$ (indicated by the subscript $d_t$). The disturbances $d_t$ should be considered as the random unobservable shocks to the players' costs, similar to the ones discussed in \cite{fudenberg1998theory}. However, in contrast to \cite{fudenberg1998theory}, the disturbances are not independent for each player but jointly affect the costs of all the players and are picked from the set $\{1,\cdots,D\}$. Each player $i \in \{1, \cdots, N\}$ has a randomized policy $x^i_{t} \in \Delta(U^i)$, based on which the control action $u^i_t$ is chosen by the player $i$. Let $x_t \in \Delta(U)$ denote the empirical joint distribution of play (given $h^i_t$) and is defined as follows. 
\begin{equation}\label{eq:empirical_joint_distribution}
    x_t(u) := \frac{1}{t}\mid \{ \tau \leq t: u_\tau = u \} \mid.
\end{equation}
Note that the empirical joint distribution is unknown to all the players, as players do not observe the control actions taken by other players. As we consider perturbed games where the costs are affected by the disturbances, we consider \emph{perturbed conditional deviations} which are an extension of conditional deviations defined in \cite{hart2000simple} and are defined for costs perturbed by disturbance $d_t \in \{1, \cdots, D\}$ at time instant $t$. The possible perturbed conditional deviation $W^i(a,b)(u_t,d_t)$ for player $i$ is defined as,
\begin{equation}\label{eq:Deviation}
    W^i(a,b)(u_t,d_t) = \begin{cases} c^i_{d_t}(b,u^{-i}_t), &\text{ if } u^i_t = a \\
        c^i_{d_t}(u^{i}_t,u^{-i}_t), &\text{ otherwise. }
        \end{cases}
\end{equation}
In contrast to the standard definition of conditional regrets in \cite{hart2000simple}, we consider \emph{perturbed conditional regrets} that are defined for costs perturbed by disturbance $d_t \in \{1, \cdots, D\}$ at time instant $t$. Since all players fix the history of play, the variables $u_t$ and $d_t$ are fixed in \eqref{eq:Deviation} for a given history of play and therefore, we have dropped them for notational simplicity for the rest of the paper. The perturbed conditional regret $CR^i_t$ for using action $a \in U^i$ up to time $t$ is defined as
\begin{multline}\label{eq:condition_regret0}
    CR^i_t(a,b) = \frac{1}{t}\bigg[\sum\limits_{\tau = 1}^t c^i_{d_\tau}(u^{i}_\tau,u^{-i}_\tau) - \sum\limits_{\tau = 1}^t W^i(a,b)\bigg], \\ \text{ with respect to any action } b \in U^i.
\end{multline}
Substituting \eqref{eq:Deviation} in \eqref{eq:condition_regret0} results in
\begin{multline}\label{eq:condition_regret}
    CR^i_t(a,b) = \frac{1}{t}\sum\limits_{\tau = 1: u^i_\tau = a}^t \big[c^i_{d_\tau}(u^{i}_\tau,u^{-i}_\tau) - c^i_{d_\tau}(b,u^{-i}_\tau)\big], \\ \text{ with respect to any action } b \in U^i.
\end{multline}
For the rest of the paper, we consider only the positive part of the perturbed conditional regrets i.e. $CR^i_{t_+} = \max\{CR^i_{t},0\}$. The following proposition implies that if every player $i$ minimizes the positive part of its perturbed conditional regret then the empirical joint distribution of play will approach the set of $\mathcal{RCE}$. 
\begin{prop} \label{th:convergence_implies_RCE}
    Define $\epsilon \geq 0$ and suppose at time $t = 1, 2, \cdots$ each player observes their history $h^i_t$ then the corresponding perturbed conditional regret $\lim\sup_{t\to \infty}CR^i_{t_+}(a,b) \leq \epsilon$ for every player $i \in \{1, \cdots, N\}$ and for every control action pair $a,b \in U^i$ (where $a \neq b$) if and only if the empirical joint distribution of play $x_t$ given by \eqref{eq:empirical_joint_distribution} converges to the set of $\mathcal{RCE}$ defined by \eqref{eq:RCE_set}.
\end{prop}
    \begin{pf}
        Consider the definition of perturbed conditional regret given by \eqref{eq:condition_regret} along with the definition of empirical joint distribution \eqref{eq:empirical_joint_distribution}. Combining them results in
        \begin{multline}\label{eq:empirical_RCE}   
            CR^i_{t_+}(a,b) = \sum\limits_{u \in U: u^i = a} x_t(u) \big[c^i_{d_t}(u^{i},u^{-i}) - c^i_{d_t}(b,u^{-i})\big], \\ \text{ with respect to any action } b \in U^i.
        \end{multline}
        Consider the definition of $\mathcal{RCE}$ in \eqref{eq:Robust_Correlated_Equilibrium} and for every player $i \in \{1, \cdots, N\}$ and for every control action pair $a,b \in U^i$ (with $a \neq b$), then $x_t(u) \to \Psi \in \mathcal{RCE}$, if $\lim\sup_{t\to \infty}CR^i_{t_+}(a,b) \leq \epsilon$ for every player $i \in \{1, \cdots, N\}$ as per the definition of $\mathcal{RCE}$. Furthermore, if we substitute $\Psi(u)$ from \eqref{eq:Robust_Correlated_Equilibrium} in place of the empirical joint distribution $x_t(u)$ in \eqref{eq:empirical_RCE} then $CR^i_{t_+}(a,b) \leq \epsilon$ and thus the converse is also true. \qed
    \end{pf}
As the regrets only make sense given the history of play \cite{perchet2013approachability}, each player is hindsight rational and that is defined as follows. 
\begin{defn}[Hindsight Rationality]
    Given the history $h^i_t$ available to player $i$ at time $t$, player $i$ updates strategy $x_{t+1}$ for $t+1$ instant such that $CR^i_{t_+} \to 0$, as $t \to \infty$ i.e. the player is internally consistent \cite{fudenberg1998theory}. 
\end{defn}
\begin{rem}
Similar to Correlated Equilibrium, the Robust Correlated Equilibrium is conditioned on the recommendation action, and therefore, a standard no-regret scheme based on external regret (i.e. comparing the sequence of actions taken against best-fixed action) may not converge to it but instead converge to a robust variant of coarse correlated equilibrium \cite{roughgarden2013cs364a}. Convergence to correlated equilibrium requires the minimization of the internal regret matrix \eqref{eq:Deviation}, \eqref{eq:condition_regret} (i.e. comparing action taken at time $t$ against best-fixed action at time $t$ for each $t$ and for the entire sequence $h^i_t$). Blackwell's Approachability theorem is used to show that the minimization of the internal regret matrix by each player implies convergence to the set of correlated equilibria in Theorem A of \cite{hart2000simple}.
\end{rem}
\subsection{Algorithm}\label{subsection:Algorithm}
 In this subsection, we present a modification of the regret matching algorithm \cite{hart2000simple} with perturbed regrets. \textcolor{black}{The idea behind regret matching in \cite{hart2000simple} is that if each player updates their mixed strategy in proportion to the \emph{regret}, then the joint distribution of play asymptotically converges to the set of $\mathcal{CE}$ with probability 1.} As each player $i$ does not observe the disturbance $d_t$ or the actions $u^{-i}_t$ of the other players directly, but instead observes the costs $c^i_{d_t}$ affected by the disturbances and joint actions, we propose a simple modification of standard time-averaged regrets in \eqref{eq:SA_CR}. The idea is to reduce variance in regrets due to disturbances, as the resulting update is an exponential moving average of past regrets. This smoothens out the effect of perturbations. Thereafter, we apply a regret matching scheme (defined later in this section and in Algorithm \ref{alg:SAbR}) to the modified perturbed conditional regrets. Convergence analysis of the algorithm shows that our proposed scheme converges to the set of $\mathcal{RCE}$. We now introduce the \emph{modified perturbed conditional regrets} $\widehat{CR}^i_{{t}_+}$ as follows, 
 \begin{multline}\label{eq:SA_CR}
        \widehat{CR}^i_{{t}_+}(a,b) = \left(1 - \frac{1}{t}\right)\widehat{CR}^i_{{t-1}_+}(a,b) \\ + \frac{1}{t}(c^i_{d_t}(a,u^{-i}_t) - c^i_{d_t}(b,u^{-i}_t)) \\
                                            + \frac{1}{t}(\widehat{CR}^i_{{t-1}_+}(a,b) - \widehat{CR}^i_{{t-2}_+}(a,b)),
    \end{multline}
with $a$ being the control action taken at time $t-1$. 
The motivation behind adding the term $\widehat{CR}^i_{{t-1}_+} - \widehat{CR}^i_{{t-2}_+}$ is to stabilize the perturbed conditional regrets despite perturbations from $d$. Note that, $\widehat{CR}^i_{{t}_+} \to 0$ implies ${CR}^i_{{t}_+} \to 0$, since we only consider positive parts of regrets.   
Furthermore, all the players will simultaneously update their control strategy at each time instant by applying Regret Matching on the modified perturbed conditional regrets. The regret matching scheme is similar to the one proposed in \cite{hart2000simple} with the only difference being that the regret matching is applied on $\widehat{CR}^i_{{t}_+}$ instead of standard regrets in \cite{hart2000simple}. The regret matching scheme builds a stochastic matrix (i.e. row entries sum up to $1$) of size $\mid U^i \mid \times \mid U^i \mid$ that consists of normalized regrets and at each time picks the row corresponding to the action taken at the previous time instant. The row picked is the mixed strategy for the next time instant. Algorithm \ref{alg:SAbR} summarizes this procedure.
\begin{algorithm}
    \caption{Perturbed Conditional Regret Matching}
    \label{alg:SAbR}
    \begin{algorithmic}[1] 
    \State \textbf{Input}: Control space $U^i$, initial strategy $x^i_1$ is uniform distribution over $U^i$.   

    \State Obtain $u^i_t \sim x^i_t$ and get corresponding $c^i_{d_t}(u^i_t,u^{-i}_t)$

    \State Let $u^i_t = a$

    \State Set $\alpha_t = \frac{1}{t}$  
    
    \For{$u^i = 1, \cdots, b, \cdots, \mid U^i \mid$}

    \State Define the vector $q_\pi \in \Delta({U^i})$ as
    \begin{equation*}
        q_\pi(b) = \begin{cases} 1, &\text{ if } a = b, \\
            0, &\text{ otherwise }.
        \end{cases}
    \end{equation*}
    
    \State Update modified perturbed regrets $\widehat{CR}^i_{{t}_+}$ \begin{multline*}
        \quad \widehat{CR}^i_{{t}_+}(a,b) = (1-\alpha_t)\widehat{CR}^i_{{t-1}_+}(a,b) \\ + \alpha_{t}(c^i_{d_t}(a,u^{-i}_t) - c^i_{d_t}(b,u^{-i}_t)) \\
                                            + \alpha_{t}(\widehat{CR}^i_{{t-1}_+}(a,b) - \widehat{CR}^i_{{t-2}_+}(a,b))
    \end{multline*}
    
    \EndFor

    \State Define normalizing constant $\mu = \sum\limits_{b \neq a} \widehat{CR}^i_{{t}_+}(a,b)$
    
    \For{$u^i = 1, \cdots, b, \cdots, \mid U^i \mid$}
    
    \State Update entries of the transition matrix $\pi^i_{{t}}(a,b)$
    \begin{equation*}\label{eq:SA_CR_RegretMatching}
        \quad \pi^i_{{t}}(a,b) = \begin{cases} \frac{1}{\mu}\widehat{CR}^i_{{t}_+}(a,b), &\text{ if } a \neq b, \\
        1 - \sum\limits_{b' \neq a}\frac{1}{\mu}\widehat{CR}^i_{{t}_+}(a,b'), &\text{ if } a = b, \\
        \frac{1}{\mid U^i \mid}, &\text{ if } \mu = 0
        \end{cases}
    \end{equation*}

    \State Update strategy $x^i_{t+1}(b) = q_\pi(a)\pi^i_{{t}}(a,b)$
    
    \EndFor
    
    \end{algorithmic}
\end{algorithm}

\subsection{Convergence analysis of Algorithm \ref{alg:SAbR}}
In this subsection, we shall analyze the convergence of Algorithm \ref{alg:SAbR}.\textcolor{black}{ We begin by stating the concepts of \emph{Enforceability} and \emph{Approachability}, followed by the statement of standard Blackwell's Approachability Theorem. }
\begin{defn}[Enforceability]
    \textcolor{black}{ Consider a $2$-player zero-sum game with scalar payoff $\mathcal{P}: R \times S \to \mathbb{R}$ where player $1$ is the minimizer. A payoff $p$ (paid by player $1$ to player $2$) is \emph{$p$-Enforceable} by player $1$, if player $1$ has a strategy $q \in \Delta(R)$ such that for any $s \in S$, player $1$ can ensure a payoff of $\mathcal{P}(q,s) \leq p$, where $\mathcal{P}(q,s) = \sum_{r \in R}q(r)\mathcal{P}(r,s)$. }
\end{defn}
\textcolor{black}{ Note that \emph{$p$-Enforceability} can be achieved if $p$ is the value of the saddle point of the game (and the player uses minimax optimal strategy) \cite{hart2013simple,von1944theory}. Now, consider a $2$-player Zero-sum game with \textbf{vector payoffs}. Let Player 1 denoted by P1 be the minimizer with control actions in finite set $R$ and Player 2 denoted by P2 be the maximizer with control actions in finite set $S$. The payoff of this game is denoted by the vector map $\mathcal{P}: R \times S \to \mathds R^D$. Each component $\mathcal{P}_d(r,s)$ of $\mathcal{P}(r,s)$ is determined by the joint actions $(r,s) \in R \times S$ chosen by P1 and P2 respectively. Let us consider a closed convex set $\mathcal{A}$. Define the support function $w_{\mathcal{A}}: \mathds R^D \to \mathds R \cup \{-\infty, \infty\}$ as,
        \begin{equation}\label{eq:Support}
            \Lambda \mapsto w_{\mathcal{A}}(\Lambda) := \sup\{\Lambda \cdot y: y \in \mathcal{A}\}.
        \end{equation}
Define $\tau \in \mathbb{N}$ and $\mathcal{P}^\tau := \mathcal{P}(r_\tau, s_\tau)$, where $(r_\tau, s_\tau)$ are the control actions chosen by players at time $\tau$. }
\begin{defn}[Approachability]
     \textcolor{black}{ A set $\mathcal{A} \subset \mathds R^{D}$ is \emph{Approachable} by P1, if P1 can guarantee the time average of payoff vector $\Bar{\mathcal{P}}: = (1/t)\sum_{\tau \leq t}\mathcal{P}^\tau$ approaches $\mathcal{A}$ as $t \to \infty$ with probability 1. }
\end{defn} 
\textcolor{black}{ We shall now state Blackwell's Approachability theorem that was used in \cite{hart2000simple} for designing the standard Regret Matching Algorithm.} 
\begin{thm}[Blackwell's Theorem \cite{blackwell1956analog}] \label{thm:Blackwell}
        \textcolor{black}{ Let $\mathcal{A} \subset \mathds R^{D}$ be a closed convex set with support function $w_{\mathcal{A}}$. The set $\mathcal{A}$ is \emph{Approachable} by P1 if and only if
        \begin{itemize}
            \item for every $\Lambda \in \mathds R^{D}$ there exists a mixed strategy $q_\Lambda \in \Delta(R)$ such that
    \begin{equation}\label{eq:Blackwell_condition}
            \Lambda \cdot \mathcal{P}(q_\Lambda, s) \leq w_{\mathcal{A}}(\Lambda) \quad \text{for all } s \in S,
        \end{equation}
         \begin{equation*}
             \text{where  }\mathcal{P}(q_\Lambda, s) = \mathds{E}_{q_\Lambda}[\mathcal{P}(r, s)] = \sum\limits_{r \in R}q_\Lambda(r)\mathcal{P}(r, s),
         \end{equation*} with summation being taken component-wise for each component $\mathcal{P}_d(r,s)$ of the vector map $\mathcal{P}$.
         \item At time $\tau+1$, P1 plays the strategy $q_\Lambda(\tau)$ (where $q_\Lambda(\tau)$ is the strategy corresponding to \eqref{eq:Blackwell_condition} for payoff $\mathcal{P}^\tau$) if $\Bar{\mathcal{P}} \notin \mathcal{A}$ and plays arbitrarily if $\Bar{\mathcal{P}}\in \mathcal{A}$. 
        \end{itemize}}
    \end{thm}
\textcolor{black}{ The Blackwell condition \eqref{eq:Blackwell_condition} says that if there exists a strategy $q_\Lambda$ such that for a static game with scalarized payoffs $\Lambda \cdot \mathcal{P}(q_\Lambda, s)$, the payoff $w_{\mathcal{A}}(\Lambda)$ is \emph{$w_{\mathcal{A}}(\Lambda)$-Enforceable} by the player for any $\Lambda$, than the set $\mathcal{A}$ is \emph{Approachable}. }
\par \textcolor{black}{Unlike standard works on internal regret such as \cite{hart2000simple,cesa2006prediction}, where the authors consider the time-average of internal regrets, the modified regrets $\widehat{CR}^i_{{t}_+}$ considered in this work are not just an time-average of internal regrets. This is explicitly shown in the equations \eqref{eq:y_t+1_full}, \eqref{eq:y_t+1} where $y$ represents $\widehat{CR}^i_{{t}_+}$ (simplified notation). Thus, we need to modify Blackwell's Approachability theorem to handle iterates that are not just time-average, and this is the focus of this subsection.} The subsection is divided into two parts where in the first part we propose a modification of Blackwell's Approachability theorem and in the second part we apply the extended Approachability theorem to derive transition matrix $\pi(a,b)$ and prove that the strategy update $x^i$ results in convergence of $\widehat{CR}^i_{{t}_+} \to 0$. By Proposition \ref{th:convergence_implies_RCE}, we know that if the perturbed conditional regrets converge to $0$ then the empirical joint distribution of play converges to the set of $\mathcal{RCE}$ (provided every player uses a strategy which ensures that their regrets converge to $0$) and thus our goal is to show that the modified perturbed conditional regrets $\widehat{CR}^i_{{t}_+} \to 0$. Note that we cannot directly use the results from \cite{hart2000simple,cesa2006prediction} as the modified perturbed conditional regrets $\widehat{CR}^i_{{t}_+}$ also consider regret due to deviation from previous estimates. \textcolor{black}{Specifically in \cite{hart2000simple,cesa2006prediction} the standard Blackwell's Approachability Theorem is used as the losses (or regrets) are time-averaged, whereas in \eqref{eq:SA_CR} the dynamics of $\widehat{CR}^i_{{t}_+}$ comprises not only time-averaged losses but instead, a combination of the previous iterates of the algorithm along with the time-averaged losses. Therefore, we modify Blackwell's Approachability Theorem for the dynamics \eqref{eq:SA_CR} instead of time-averaged dynamics in standard Blackwell's Approachability Theorem (Theorem \ref{thm:Blackwell}).}  
\subsubsection{Modifying Blackwell's Approachability theorem} 
Consider a $2$-player perturbed Zero-Sum finite repeated game $\Gamma'_2 = ( 2, U^1, U^2, l, D)$ with losses $l: U^1 \times U^2 \times D \to \mathds R^L$ (with $L$ being the arbitrary dimension of the loss vector) incurred by Player 1. Let $l_{t+1} = l(a_{t+1},b_{t+1},d_{t+1})$ represent the bounded loss incurred by player $i$ at time $t+1$ if the players choose actions $(a_{t+1},b_{t+1}) \in U^1 \times U^2$ with the losses being perturbed by disturbance $d_{t+1} \in \{1, \cdots, D\}$. Consider the following update equation,
\begin{align}\label{eq:y_t+1_full}
    y_{t+1} = \left(1-\frac{1}{t+1}\right)y_{t} + \frac{1}{t+1}l_{t+1} + \frac{1}{t+1}\left( y_{t} - y_{t-1} \right),
\end{align}
where $t \in \mathbb{N}$, $y_1 = l_{1}$, $y_0 = 0$, and we change the time index to $t+1$ since $y_{-1}$ is undefined. 
\begin{defn}[Approachability of $y_t$]\label{defn:Approachability}
     The set $\mathcal{A} \subset \mathds R^L$ is Approachable by player $1$ if the player $1$ has a randomized strategy such that no matter how player $2$ plays, 
    \begin{equation}\label{eq:Approachability}
        \lim\limits_{t \to \infty} \text{dist}\bigg( y_t, \mathcal{A} \bigg) = 0, \quad \text{almost surely,}
    \end{equation}
    where $\text{dist}(\mathbf{j},\mathcal{A}) = \inf_{\mathbf{k} \in \mathcal{A}} \norm{\mathbf{j} - \mathbf{k}}$ is the Euclidean distance of $\mathbf{j}$ from the set $\mathcal{A}$. 
\end{defn} 
Note that Approachability of $y_t$ implies Approachability of ${CR}^i_{{t}_+}$ as the form of $y_t$ is same as \eqref{eq:SA_CR} and convergence of $y_t$ implies that the term $y_t - y_{t-1} \approx 0$ and observe that in this case \eqref{eq:y_t+1_full} is approximately equivalent to 
\begin{equation*}
    y_{t+1} \approx \left(1-\frac{1}{t+1}\right)y_{t} + \frac{1}{t+1}l_{t+1},
\end{equation*}
which is same as the positive part of \eqref{eq:condition_regret} (with time index $t$ instead of $t+1$). The following theorem generalizes Blackwell's Approachability theorem for iterates of the form \eqref{eq:y_t+1_full}. 
\begin{thm}\label{th:Extended_Blackwell_Approachability}
    Consider the game $\Gamma'_2$ and let $y_t$ be updated by Player $1$ as per \eqref{eq:y_t+1_full}. The iterate $y_t$ \emph{Approaches} to the set $\mathcal{A}$ as per \eqref{eq:Approachability} if and only if the following conditions are satisfied: \begin{itemize}
        \item For every $\Lambda \in \mathds R^L$ there exists a strategy $q_\Lambda \in \Delta(U^1)$ for Player $1$ such that
        \begin{equation}\label{eq:Blackwell_like_condition1}
            \Lambda \cdot l(q_\Lambda,b,d) \leq w_{\mathcal{A}}(\Lambda), \text{ } \forall b \in U^2, d \in \{1, \cdots, {D}\},
        \end{equation}
        where $l(q_\Lambda,b,d) = \mathds{E}_{q_\Lambda}[l(a,b,d)] = \sum\limits_{a \in U^1}q_\Lambda(a)l(a,b,d)$.
        \item At time $t+1$, player $1$ plays the strategy $q_\Lambda(t)$, where $q_\Lambda(t)$ is the strategy corresponding to \eqref{eq:Blackwell_like_condition1} for $y_t$ if $y_t \notin \mathcal{A}$, and plays arbitrarily if $y_t \in \mathcal{A}$. 
    \end{itemize}
\end{thm}
\begin{pf}
    Let the loss incurred at time instant $t$ be denoted by $l_t$. Since the losses are bounded at each time instant, $l_t$ can be normalized by dividing with the maximum possible loss such that $\norm{l_t} \leq 1$ without loss of generality. Let $P(y_t)$ denote the projection of $y_t$ on the set $\mathcal{A}$,
    \begin{equation*}\label{eq:Project_defn}
        P(y_t) = \arg\min\limits_{z \in \mathcal{A}} \norm{y_t - z}. 
    \end{equation*}
    Since $\mathcal{A}$ is a convex set, the projection $P(y_t)$ is a unique point on $\mathcal{A}$ \cite{boyd2004convex}. Define $\Lambda_{t-1}$ as the unit vector pointing in the direction of the set $\mathcal{A}$,
    \begin{equation}\label{eq:lambda_unitVector}
        \Lambda_{t-1} := \frac{y_{t-1} - P(y_{t-1})}{\norm{y_{t-1} - P(y_{t-1})}}.
    \end{equation}
    The squared distance $\text{dist}^2(y_t,\mathcal{A})$ is calculated as,
    \begin{eqnarray}\label{eq:CR_proof1}
        \text{dist}^2(y_t,\mathcal{A}) = \norm{y_{t} - P(y_{t})}^2 \leq \norm{y_{t} - P(y_{t-1})}^2, 
    \end{eqnarray}
    where the inequality is due to the optimality of the projection operator for each time step. Substituting \eqref{eq:y_t+1_full} in the last term of the above equation results in,    
    \begin{align}\label{eq:CR_proof2}
    \begin{split}
        &\norm{y_{t} - P(y_{t-1})}^2  = \\ &\norm{ \frac{t-1}{t}y_{t-1} + \frac{l_t(a,b)}{t} + \frac{1}{t}(y_{t-1} - y_{t-2}) - P(y_{t-1})}^2 \\ & =
        \bigg\lVert\frac{t-1}{t}\left(y_{t-1} - P(y_{t-1})\right) + \frac{l_t(a,b) - P(y_{t-1})}{t} \\ &+ \frac{1}{t}(y_{t-1} - y_{t-2}) \bigg\rvert\bigg\rvert^2 = \\ &
        \norm{\frac{t-1}{t}\left(y_{t-1} - P(y_{t-1})\right) +  \frac{l_t(a,b) - P(y_{t-1})}{t}}^2 + \\ & \frac{1}{t^2}\norm{(y_{t-1} - y_{t-2})}^2 +  \frac{2}{t^2}\bigg[ (t-1)(y_{t-1} - P(y_{t-1})) \\ & + l_t(a,b) - P(y_{t-1}) \bigg]\cdot\left[y_{t-1} -  y_{t-2}\right].
    \end{split}    
    \end{align}
    The first term in the above equation (in the last step) can be expanded as follows 
    \begin{eqnarray*}
        &\norm{\frac{t-1}{t}\left(y_{t-1} - P(y_{t-1})\right) +  \frac{l_t(a,b) - P(y_{t-1})}{t}}^2 = \\&
        \left(\frac{t-1}{t}\right)^2\norm{y_{t-1} - P(y_{t-1})}^2 + \frac{1}{t^2}\lvert\lvert l_t(a,b) - P(y_{t-1})\\&\rvert\rvert^2  
        + 2\frac{t-1}{t^2}\left[y_{t-1} - P(y_{t-1})\right]\cdot\left[l_t(a,b) - P(y_{t-1})\right]
    \end{eqnarray*}
    Remaining terms of the equation \eqref{eq:CR_proof2} are simplified as,
    \begin{eqnarray*}
        &\frac{1}{t^2}\norm{(y_{t-1} - y_{t-2})}^2 +  \frac{2}{t^2}\bigg[ (t-1)(y_{t-1} - P(y_{t-1})) \\& + l_t(a,b) - P(y_{t-1}) \bigg]\cdot\left[y_{t-1} -  y_{t-2}\right] = \\&
        \frac{1}{t^2}\left[ y_{t-1} - y_{t-2} \right] \cdot [ y_{t-1} - y_{t-2} + 2(t-1)(y_{t-1} \\&  - P(y_{t-1})) + l_t(a,b) - P(y_{t-1})]
    \end{eqnarray*}
    Furthermore as $\norm{l_t} \leq 1$, the $\norm{l_t(a,b) - P(y_{t-1})}$ will be $\leq 2$. Thus the inequality given by \eqref{eq:CR_proof1} can be simplified by multiplying by $t^2$ and rearranging the terms as,
    \begin{multline}\label{eq:CR_proof_final1}
        t^2\norm{y_{t} - P(y_{t})}^2 - (t-1)^2\norm{y_{t-1} - P(y_{t-1})}^2 \leq\\  4 + 2(t-1)[y_{t-1} - P(y_{t-1})]\cdot[l_t(a,b)-P(y_{t-1})] + \\ \left[ y_{t-1} - y_{t-2} \right] \cdot [ y_{t-1} - y_{t-2} + 2(t-1)(y_{t-1} - P(y_{t-1})) \\ + l_t(a,b) - P(y_{t-1})].
    \end{multline}
    Summing up both the sides of the inequality \eqref{eq:CR_proof_final1} for time $t=1,\cdots, T$ results in a telescopic sum of the left-hand side of the inequality \eqref{eq:CR_proof_final1} as $T^2\norm{y_{T} - P(y_{T})}$  
    \begin{align}\label{eq:CR_proof_final2}
    \begin{split}
        &T^2\norm{y_{T} - P(y_{T})}^2 \leq \sum\limits_{t = 1}^T 4 + \sum\limits_{t = 1}^T \bigg(2(t-1)[y_{t-1} - P(y_{t-1})\\&] \cdot [l_t(a,b)-P(y_{t-1})] +  \left[ y_{t-1} - y_{t-2} \right] \cdot [ y_{t-1} - y_{t-2} \\&+ 2(t-1)(y_{t-1} - P(y_{t-1})) + l_t(a,b) - P(y_{t-1})]\bigg),
    \end{split}
    \end{align}
    and on the right-hand side the term $y_{t-1} - y_{t-2}$ telescopes to $y_{T-1}$. Denote $(t-1)\norm{y_{t-1} - P(y_{t-1})}$ as $K_{t-1}$ and dividing both sides of \eqref{eq:CR_proof_final2} by $T^2$ results in
    \begin{multline}\label{eq:CR_proof_final3}
        \norm{y_{T} - P(y_{T})}^2 \leq \\ \frac{4}{T} + \frac{2}{T}\sum\limits_{t = 1}^T K_{t-1} \Lambda_{t-1} [l_t(a,b)- P(y_{t-1})] + \\ \frac{y_{T-1}}{T}  \bigg[ \frac{y_{T-1}}{T} + \frac{2}{T}\sum\limits_{t = 1}^T \bigg( K_{t-1} \Lambda_{t-1} + l_t(a,b) - P(y_{t-1})\bigg)\bigg].
    \end{multline}
    Using the definition of the support function given by \eqref{eq:Support} and \eqref{eq:lambda_unitVector}, the condition given by \eqref{eq:Blackwell_like_condition1} can be written as
    \begin{multline}\label{eq:Blackwell_like_condition2}
        \Lambda_{t-1} \cdot l_{t}(q_{\Lambda_{t-1}},b) \leq  \Lambda_{t-1} \cdot P(y_{t-1}), \\ \forall b \in U^2, \text{ } \forall d \in \{1, \cdots, {D}\}.
    \end{multline}
    Substituting \eqref{eq:Blackwell_like_condition2} in \eqref{eq:CR_proof_final3} leads to 
    \begin{multline}\label{eq:CR_proof_final4}
        \norm{y_{T} - P(y_{T})}^2 \leq \\ \frac{4}{T} + \frac{2}{T}\sum\limits_{t = 1}^T K_{t-1} \Lambda_{t-1} [l_t(a,b)- l_t(a,q_{\Lambda_{t-1}})] + \\ \frac{y_{T-1}^2}{T^2} + \bigg[ \frac{2y_{T-1}}{T^2}\sum\limits_{t = 1}^T \bigg( K_{t-1} \Lambda_{t-1} + l_t(a,b) - P(y_{t-1})\bigg)\bigg].
    \end{multline}
    The terms $K_{t-1}$ and $l_t(a,b) - P(y_{t-1})$ are bounded between $0$ and $2$ whereas the term $l_t(a,b) - l_t(q_{\Lambda_{t}},b)$ is a Martingale Difference Sequence (see definition \ref{defn:MDS} in Appendix) with respect to the random variable $a_t$ which is sampled from the distribution $q_{\Lambda_{t-1}}$ i.e. 
    \begin{align*}
        \mathds E_{q_{\Lambda_{t-1}}} [l_t(a_t,b) - l_t(q_{\Lambda_{t}},b)\mid a_1, \cdots, a_t] = 0, 
     \end{align*}   
        holds almost surely. Convergence of remaining terms of \eqref{eq:CR_proof_final4} is studied in the following Lemma.
    \begin{lem}
        The component-wise value of $y_{T}$ generated by \eqref{eq:y_t+1_full} is bounded by $\mathds E_{q_\Lambda}[l_t(a,b)]$ almost surely.
    \end{lem}
    \begin{pf} 
    Recall that each component of the vector $y_t \in \mathds R_+$ and $y_t$ is updated as per \eqref{eq:y_t+1_full}, 
    \begin{align}\label{eq:y_t+1}
    \begin{split}
        y_{t+1} =& \left(1-\frac{1}{t+1}\right)y_{t} + \frac{1}{t+1}l_{t+1} + \frac{1}{t+1}\left( y_{t} - y_{t-1} \right),\\
            =& y_{t} - \frac{1}{t+1}y_{t-1} + \frac{1}{t+1}l_{t+1},
    \end{split}
    \end{align}
    where $t \in \mathbb{N}$, $y_1 = l_{1}$, $y_0 = 0$. The outline of the proof is as follows. For the remaining proof, all the analysis will be component-wise for the vector $y_t$. The analysis of sequence $(y_t)$ generated by \eqref{eq:y_t+1} is separated into analysis of the sequence generated by an autonomous component of \eqref{eq:y_t+1} denoted by $\Tilde{y}_{t+1}$ and the sequence generated by \eqref{eq:y_t+1} with input loss $l_{t+1}$. Firstly, we consider the sequence generated by the autonomous component of the equation \eqref{eq:y_t+1} i.e. \eqref{eq:y_t+1} without input loss $l_{t+1}$ (defined more precisely in \eqref{eq:autonomous_eqn}). We observe that the sequence $(\Tilde{y}_t)$ is decreasing and bounded from below by $0$, so it is convergent. Then we prove by contradiction that its limit is $0$. Thereafter, we show that the sequence $(y_t)$ generated by the non-autonomous system will be bounded by $\mathds{E}_{q_\Lambda}[l_{t+1}]$. 
    Consider the autonomous component of the equation \eqref{eq:y_t+1} (everything except $(1/(t+1))l_{t+1}$ in \eqref{eq:y_t+1}) as follows
    \begin{align}\label{eq:autonomous_eqn}
        y_{t+1} = y_t - \frac{1}{t+1}y_{t-1}.
    \end{align}
    We claim that the sequence generated by \eqref{eq:autonomous_eqn}, $\Tilde{y}_\tau \to 0$ for some finite time $\tau$. The proof of this claim is as follows. Suppose that the sequence generated by \eqref{eq:autonomous_eqn} has a positive limit. Specifically for any $\epsilon > 0$, assume that there exists $\tau:= \tau(\epsilon)$ such that $\Tilde{y}_\tau > \epsilon$. Than from \eqref{eq:autonomous_eqn},
    \begin{align*}
        \Tilde{y}_\tau = \Tilde{y}_{\tau-1} - \alpha_\tau \Tilde{y}_{\tau-2}, \quad \text{where } \alpha_\tau = \frac{1}{\tau}.
    \end{align*} 
    \begin{align*}
        \text{If } \Tilde{y}_\tau > \epsilon \text { than } \Tilde{y}_{\tau-1} > \epsilon + \alpha_\tau \Tilde{y}_{\tau-2}.
    \end{align*}
    Since $\Tilde{y}_\tau \in \mathds R_+$, $ \forall \tau$, we can iterate backwards the previous inequality as
    \begin{align*}
        \text{If } \Tilde{y}_{\tau-1} > \epsilon &\text { than } \Tilde{y}_{\tau-2} > \epsilon + \alpha_\tau \Tilde{y}_{\tau-3}, \\
        \cdots \quad &\quad \cdots\\
        &\quad \Tilde{y}_0 > \epsilon.
    \end{align*}
    But $\Tilde{y}_0$ cannot be greater than $\epsilon$ since we initialize \eqref{eq:y_t+1} in Algorithm \ref{alg:SAbR} with $\Tilde{y}_0 = 0$ (contradiction). Since this is true for any $\epsilon > 0$ and $\Tilde{y}_\tau \in \mathds R_+$, $ \forall \tau$, we can conclude that $\Tilde{y}_\tau \to 0$ in some finite time $\tau$. Now consider the non-autonomous system given by \eqref{eq:y_t+1} and modify the equation as follows
    \begin{multline*}
        y_{t+1} = y_{t} - \frac{1}{t+1}y_{t-1} + \frac{1}{t+1}(l_{t+1} - \mathds{E}_{q_\Lambda}[l_{t+1}]) \\ + \frac{1}{t+1}\mathds{E}_{q_\Lambda}[l_{t+1}].
    \end{multline*}
    Recall that $\mathds{E}_{q_\Lambda}[l_{t+1}(a,b)] = l(q_\Lambda,b)$ as defined in \eqref{eq:Blackwell_like_condition1}. Note that the term $l_{t+1} - \mathds{E}_{q_\Lambda}[l_{t+1}]$ is a Martingale Difference Sequence. Therefore we can conclude $y_T \leq \mathds{E}_{q_\Lambda}[l_{t+1}]$ almost surely. \qed
    \end{pf}
    Thus all the terms in \eqref{eq:CR_proof_final4} are either bounded or are a Martingale Difference Sequence. Let the Martingale Difference Sequence generated by $\norm{y_{T} - P(y_{T})}^2$ be bounded by constants $0 < k_1, k_2, \cdots k_t, \cdots < \infty$, then for any $m>0$, we have the following concentration inequality due to Azuma–Hoeffding inequality (see Theorem \ref{th:Azuma} in Appendix) we have
    \begin{align*}
        \mathds P[\norm{y_{T} - P(y_{T})}^2 > m] \leq \text{exp} \left( \frac{-2m^2}{\sum_{t=1}^T k_t^2} \right) < \infty,
    \end{align*}
    and thus by Borel-Cantelli Lemma (see Lemma \ref{lemma:Borel-Cantelli} in Appendix), we can conclude that for sufficiently large $T$, $\norm{y_{T} - P(y_{T})}^2 \to 0$ almost surely. \qed
\end{pf}
\subsubsection{Convergence of Algorithm \ref{alg:SAbR} using Theorem \ref{th:Extended_Blackwell_Approachability}}
\begin{prop}
    The modified perturbed conditional regret $\widehat{CR}^i$ and the corresponding perturbed conditional regret ${CR}^i$ in Algorithm \ref{alg:SAbR} converges to 0 for all actions $a, b \in U^i$ and for all players $i \in N$ if all the players use Algorithm \ref{alg:SAbR} in a decentralized manner.
\end{prop}
\begin{pf}
Consider the static game at time $t$ and fix the control actions $u_t = [u^i_t, u^{-i}_t]$ taken by all players, and fix the disturbance $d_t$ experienced by each of the players. Then an auxiliary stage game is defined as follows.
     Let $l^i_{t}$ denote the perturbed conditional regret incurred by player $i$ at time $t$ (i.e. regret per time instant for not choosing control action $b$ instead of control action $a$) as
    \begin{equation}\label{eq:l_t}
        l^i_{t}(u_t,d_t)(a,b) := \begin{cases} c^i_{d_t}(a,u^{-i}_t) - c^i_{d_t}(b,u^{-i}_t), &\text{ if } u^i_t = a \\
        0, &\text{ otherwise. }
        \end{cases}
    \end{equation}
Consider Theorem \ref{th:Extended_Blackwell_Approachability} with \eqref{eq:y_t+1_full}. We consider the Approachability of Auxiliary Game loss \eqref{eq:l_t} to the set $\mathcal{S}_-:=\{x \in \mathbb{R}^L: x \leq 0\}$, where $L$ is given as follows,
    \begin{equation}\label{eq:Index_Set}
        L:= \{(a,b) \in U^i \times U^i: a \neq b\}.
    \end{equation} 
Consider $y_t$ (recall that it represents $\widehat{CR}^i$ in \eqref{eq:SA_CR}) as per \eqref{eq:y_t+1_full} and note that $P(y_t) = 0$, where $P(y_t)$ is the projection of $y_t$ on $\mathcal{S}_-$. Define 
\begin{equation*}
    \pi_t := \frac{y_t}{\norm{y_t}_1}, \quad \text{ where} \norm{y_t}_1 \text{ is $1$-norm of } y_t,
\end{equation*}   
The condition \eqref{eq:Blackwell_like_condition1} can now be written as
\begin{align*}
\begin{split} 
    \pi_t(a,b) \cdot l_t^i(a,q_{\pi_t}) \leq 0 \text{ } \forall a \in U^i, \text{ } \forall d \in \{1, \cdots {D}\}, \\
    \pi_{{t}}(a,b)q_{\pi_t}(b)[c^i_{d_t}(a,u^{-i}_t) - c^i_{d_t}(b,u^{-i}_t)] \leq 0, \\ \text{ } \forall a \in U^i, \text{ } \forall d \in \{1, \cdots {D}\}.
\end{split}
\end{align*}
Choosing $q_\pi \in \Delta({U^i})$ as 
\begin{equation*}
        q_\pi(b) = \begin{cases} 1, &\text{ if } a = b, \\
            0, &\text{ otherwise },
        \end{cases}
        \forall b \in \mid U^i \mid,
\end{equation*}
results in the condition \eqref{eq:Blackwell_like_condition1} being satisfied for all $a \in U^i$ and for all $d \in \{1, \cdots {D}\}$ as an equality.
\qed
\end{pf}
\section{Simulation studies}\label{Section:Simulation}
\begin{figure}
    \centering
    \includegraphics[width=7cm]{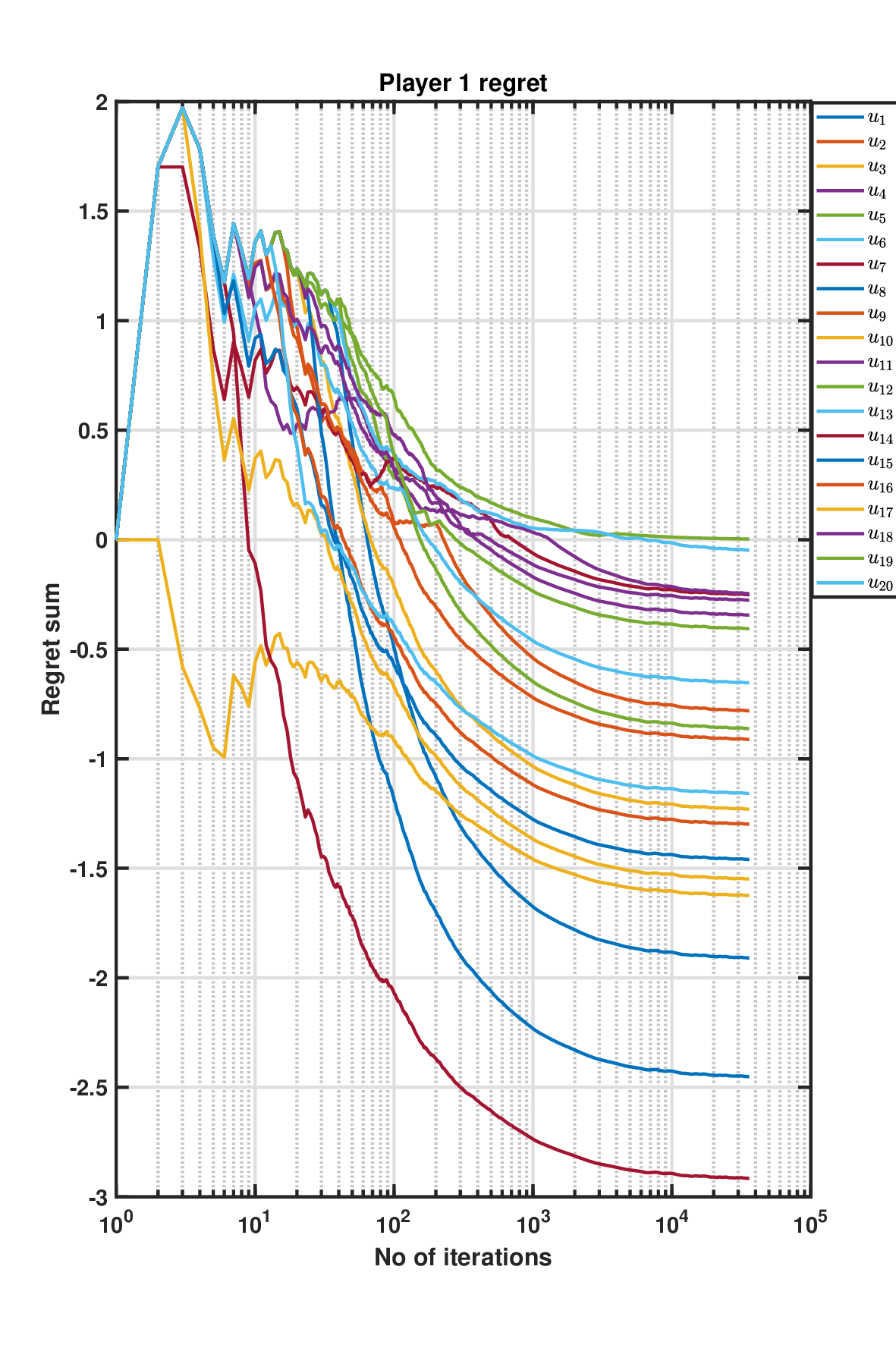}
    \caption{Regrets experienced by Player 1.}
    \label{fig:P1_regret}
\end{figure}
\begin{figure}
    \centering
    \includegraphics[width=7cm]{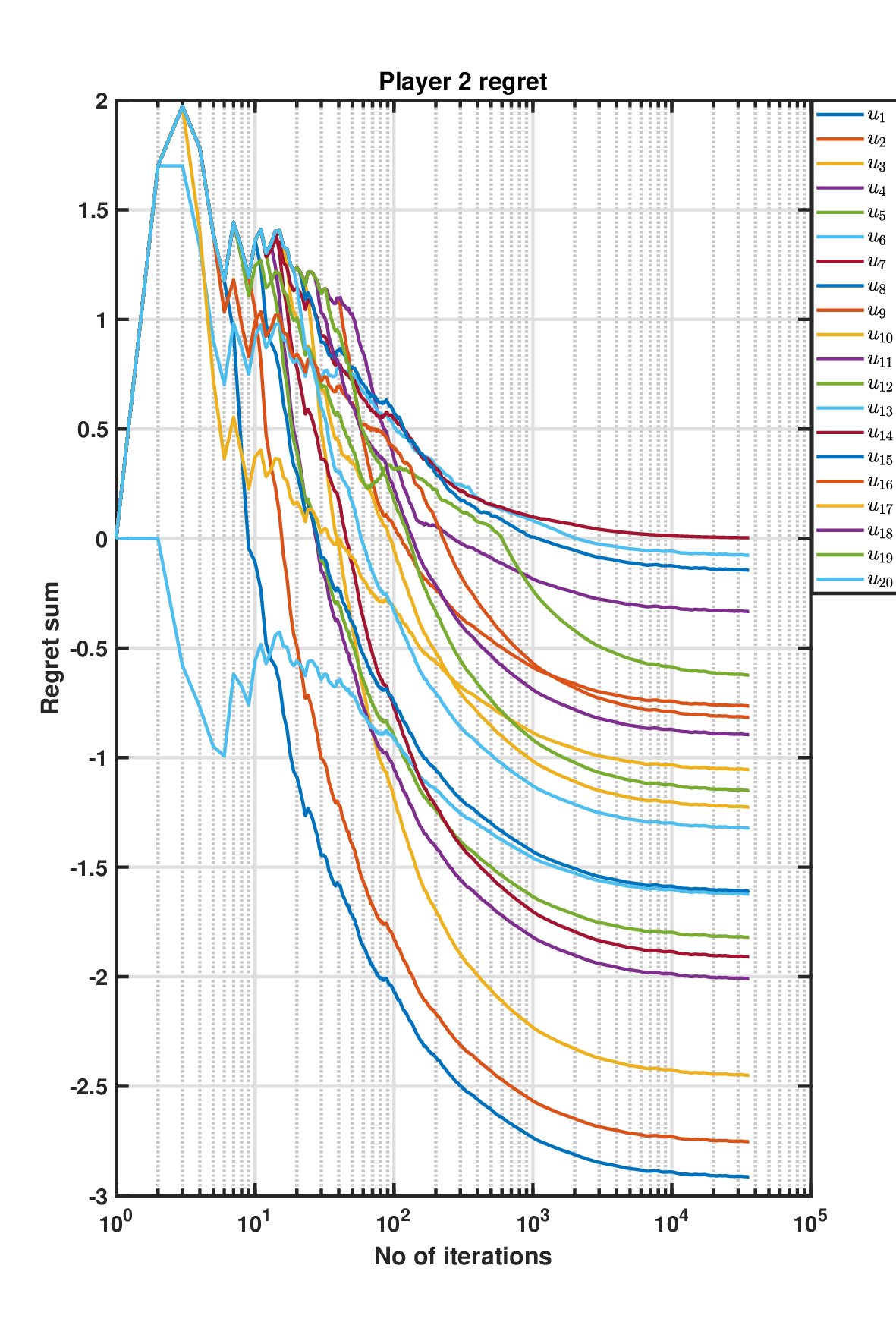}
    \caption{Regrets experienced by Player 2.}
    \label{fig:P2_regret}
\end{figure}
\begin{figure}
    \centering
    \includegraphics[width=7cm]{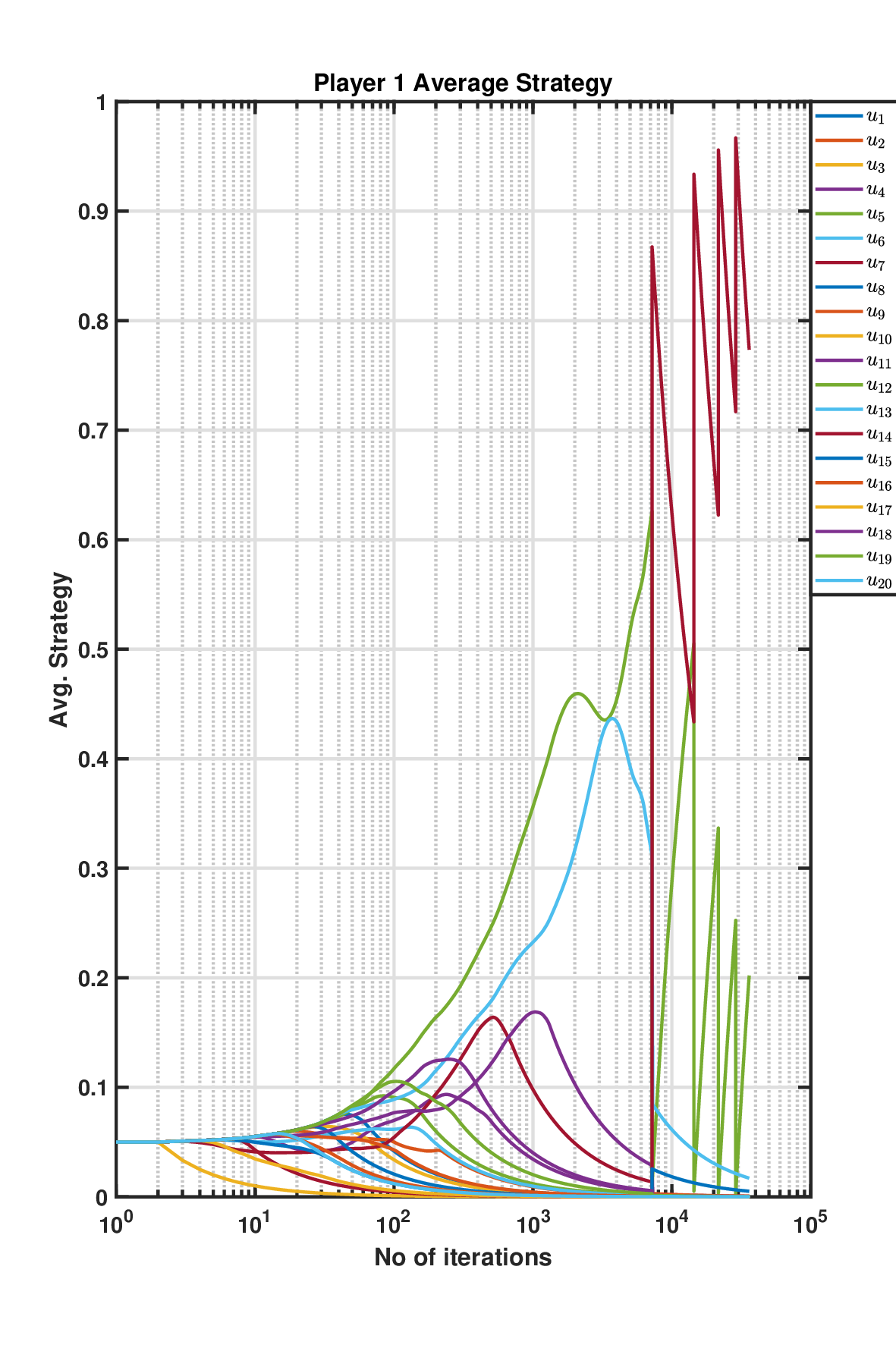}
    \caption{Averaged empirical mixed strategy for Player 1.}
    \label{fig:P1_Avg}
\end{figure}
\begin{figure}
    \centering
    \includegraphics[width=7cm]{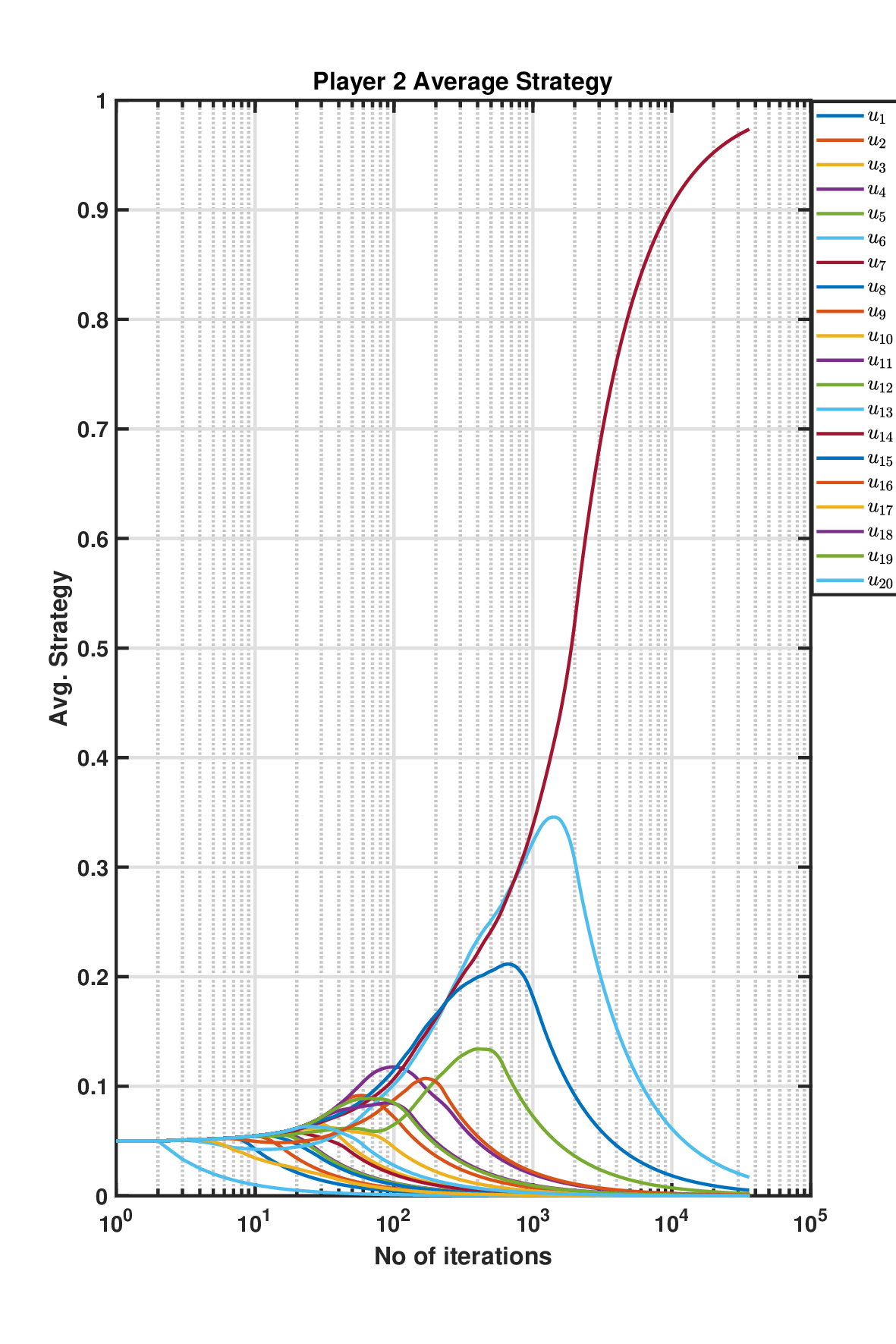}
    \caption{Averaged empirical mixed strategy for Player 2.}
    \label{fig:P2_Avg}
\end{figure}
\begin{figure}
    \centering
    \includegraphics[width=7cm]{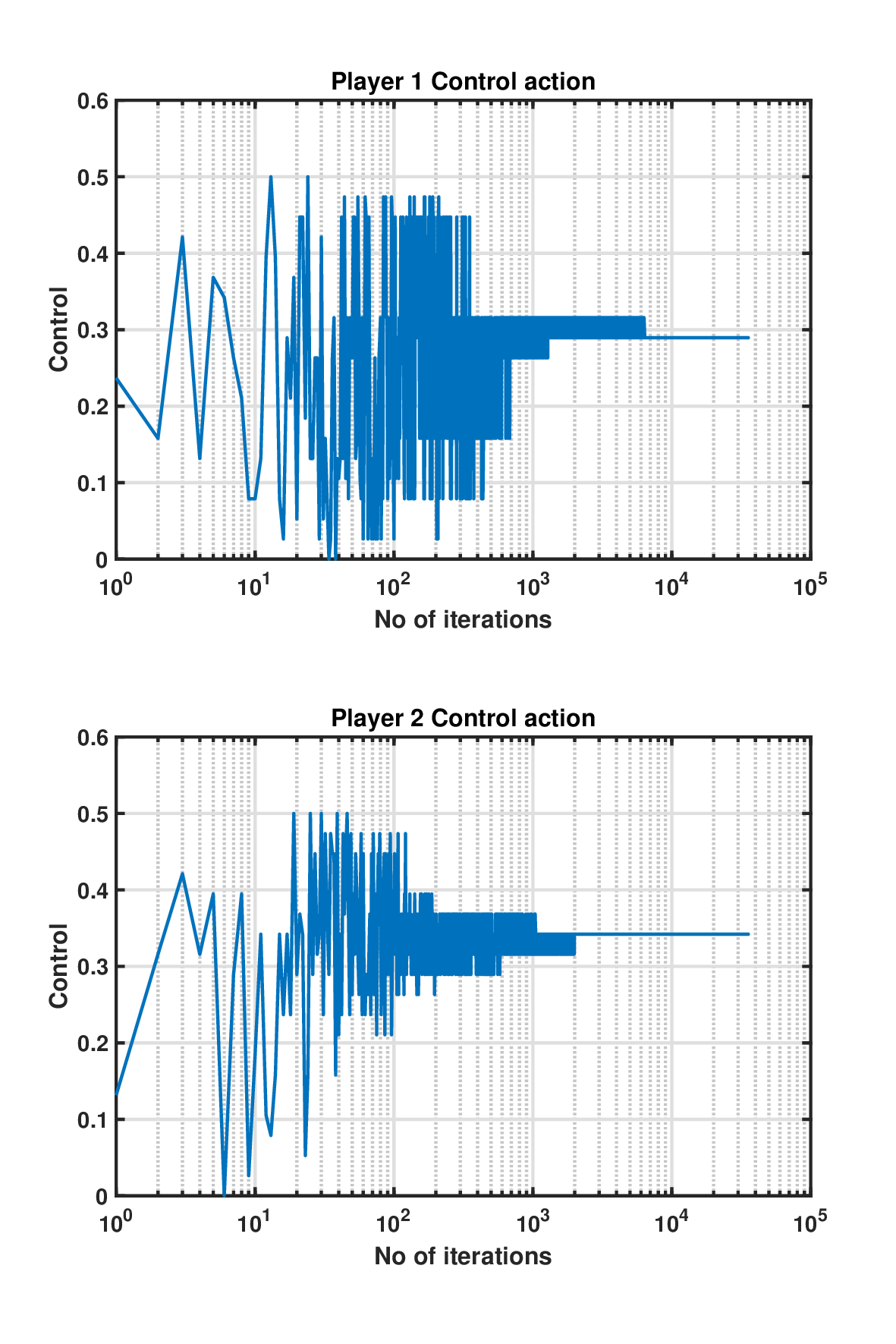}
    \caption{Realized control actions by both the players.}
    \label{fig:control}
\end{figure}
\begin{figure}
    \centering
    \includegraphics[width=7cm]{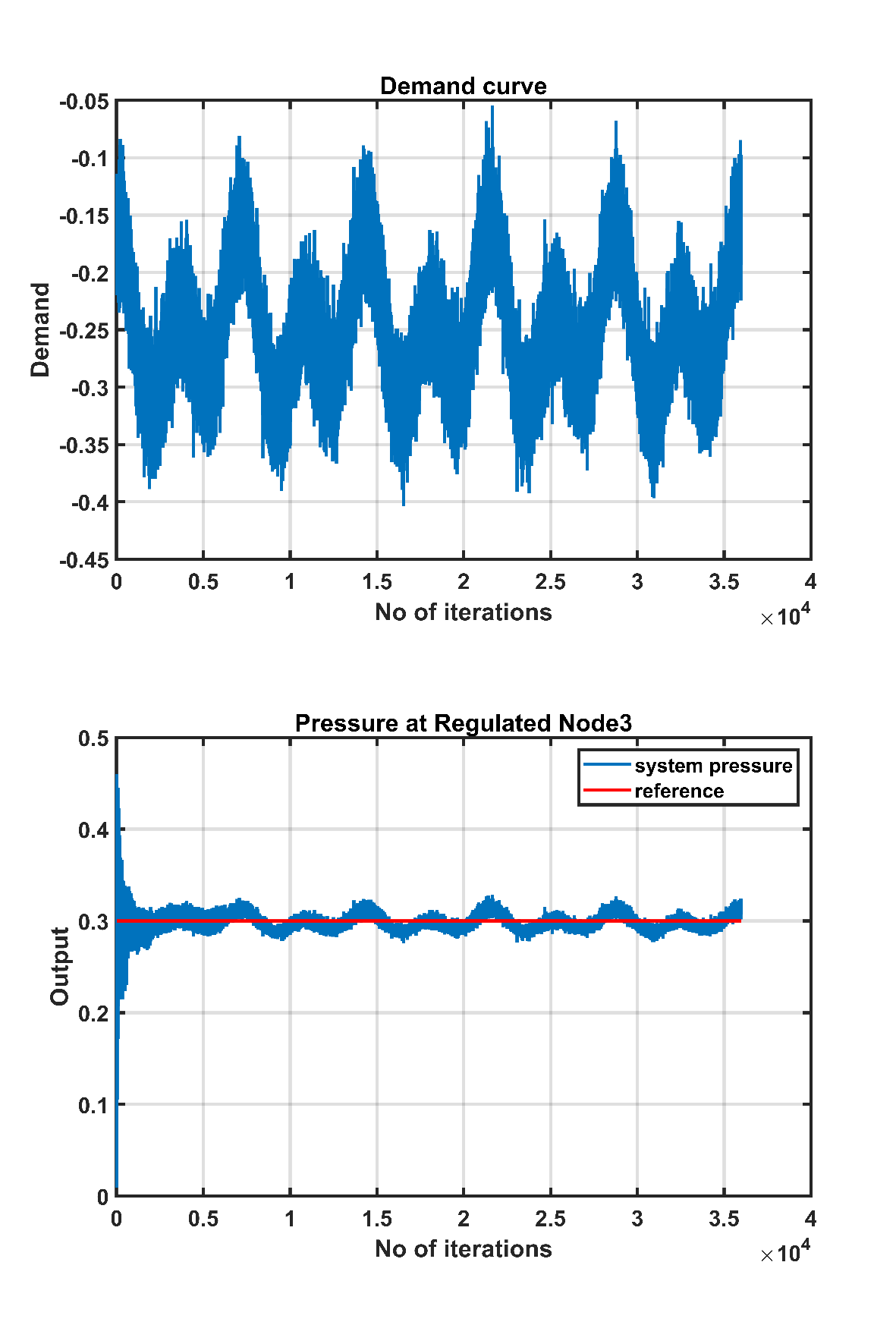}
    \caption{Pressure is regulated despite consumer disturbances.}
    \label{fig:Pressure}
\end{figure}
\begin{figure}
    \centering
    \includegraphics[width=7cm]{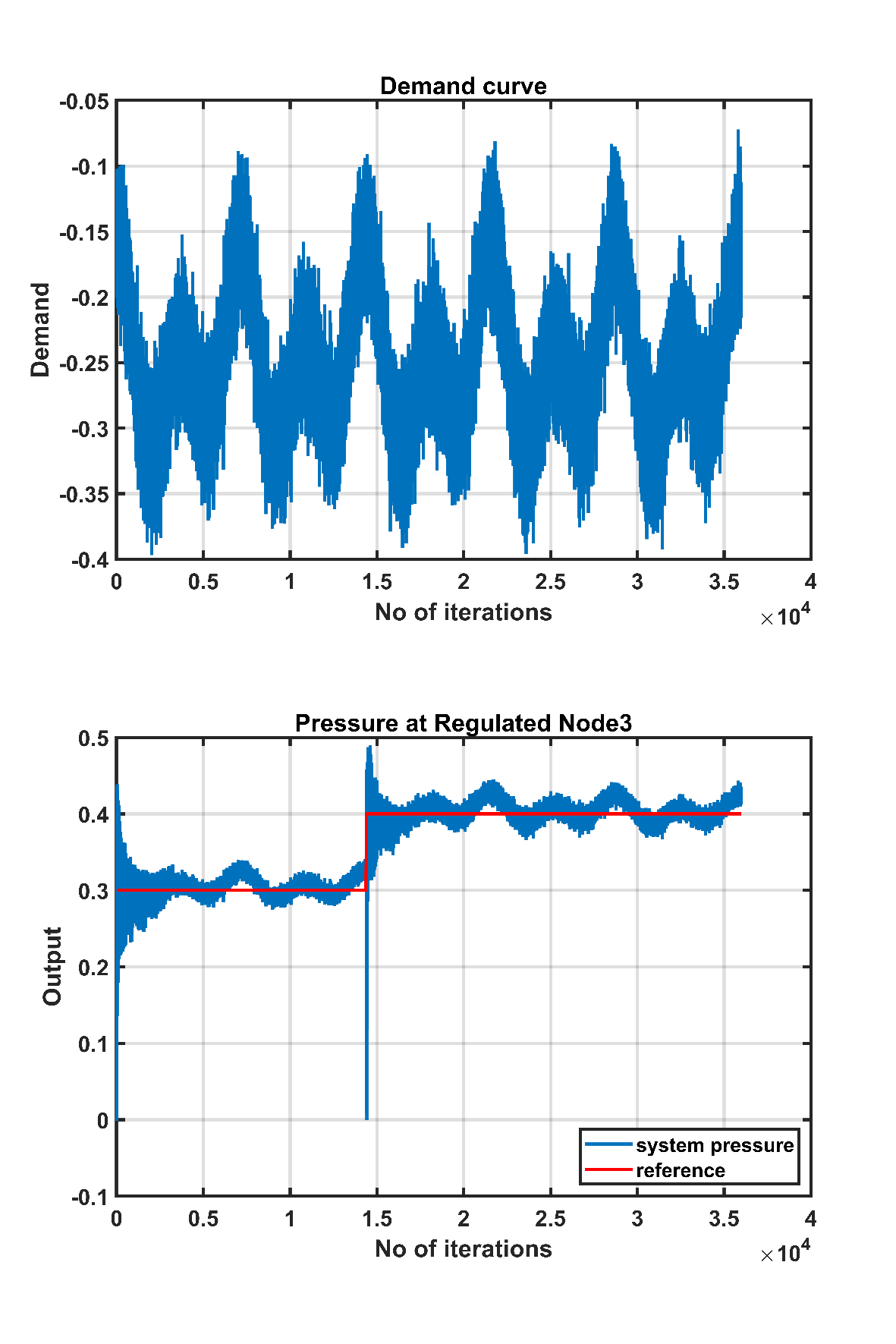}
    \caption{The players handle changing set points satisfactorily.}
    \label{fig:Pressure_SP}
\end{figure}
\begin{figure}
    \centering
    \includegraphics[width=7cm]{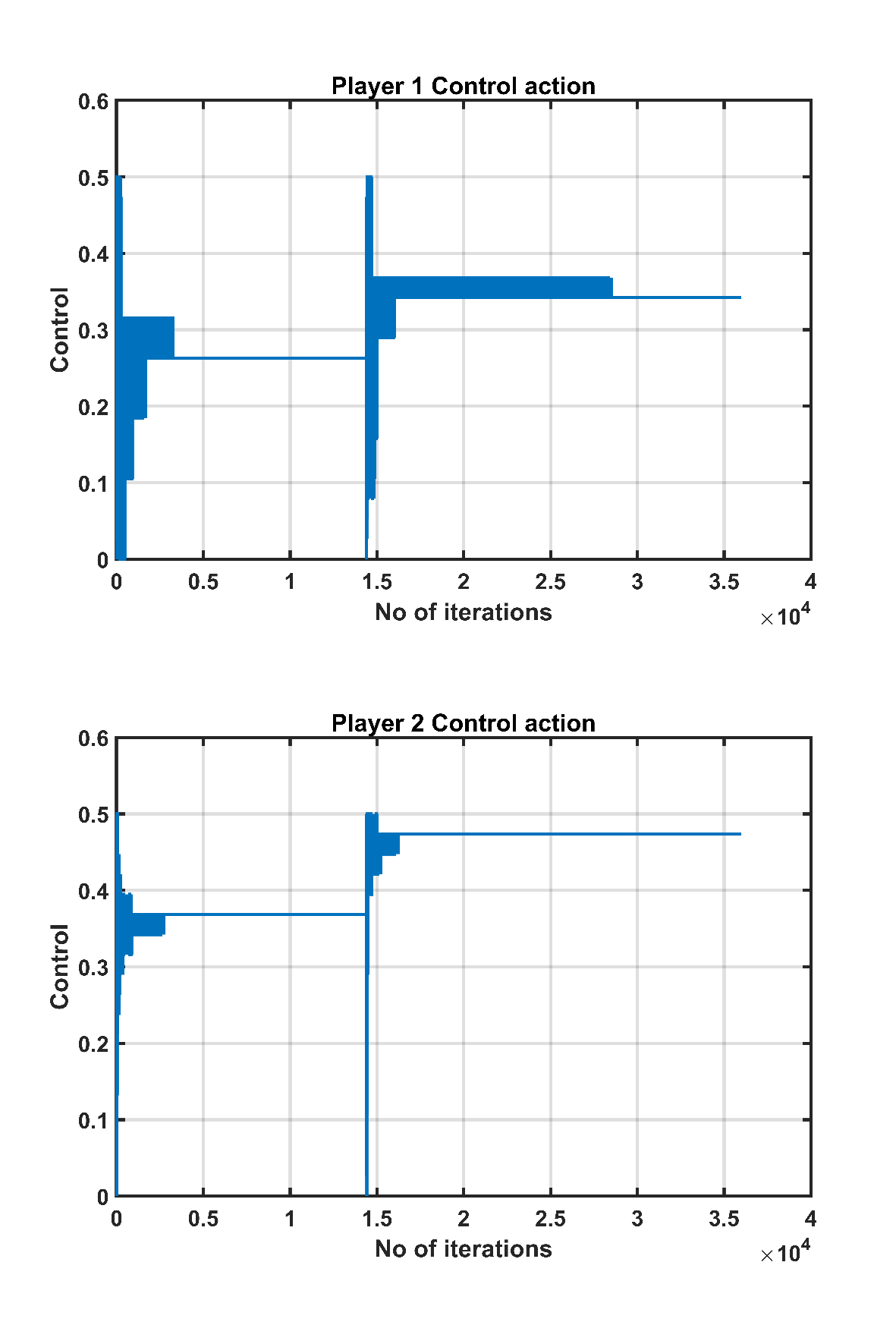}
    \caption{Realized control actions with change in the set point.}
    \label{fig:Control_SP}
\end{figure}

In this section, we apply the algorithm to a decentralized control problem for a water distribution network with pumping stations acting as controllers. The goal of each decentralized controller is to ensure tracking of the desired pressure set point in the system while minimizing its own energy consumption. The decentralized optimal control problem is described precisely in our previous work \cite{misra2023decentralized}, where we propose model-based minimax strategies for each player and implement them on a laboratory setup that emulates a real-life water distribution network.  Firstly, we present the modified perturbed regrets $\widehat{CR}^i_{{t}_+}$ calculated by each player using Algorithm \ref{alg:SAbR} on semi-logarithmic plots in Fig. \ref{fig:P1_regret} and Fig. \ref{fig:P2_regret}. The x-axis is on the logarithmic scale while the y-axis is on a regular scale. Next, we present the average of the empirical mixed strategy of the players (which jointly converge to the $\mathcal{RCE}$) and the realized control actions on semi-logarithmic plots in Fig. \ref{fig:P1_Avg}, fig. \ref{fig:P2_Avg} and Fig. \ref{fig:control}. 
Simulation results show that the controllers calculate an optimal solution in the sense of a $\mathcal{RCE}$ while being game agnostic i.e. they are unaware of the existence of the other controllers and the dynamics of the system. The results presented here are an improvement on the work presented in \cite{misra2023decentralized} as our solution is more efficient for individual players and better regulation of pressure is achieved as shown in Fig. \ref{fig:Pressure}. 
Further experiments demonstrate that our algorithm can also handle changing set points (by re-initializing if it detects a change in the set point). This is shown in the plots in Fig. \ref{fig:Pressure_SP} and Fig. \ref{fig:Control_SP}. 

\section{Conclusion and Future Work}
We have successfully extended the concept of $\mathcal{CE}$ to time-varying games by introducing $\mathcal{RCE}$ and a decentralized algorithm that each player can use to learn $\mathcal{RCE}$ in the Unkown game setting. The convergence analysis of the algorithm and simulation studies validate our approach. We conjecture that Theorem \ref{th:Extended_Blackwell_Approachability} can be extended for games with continuous but bounded disturbances as per the simulation results and a proof for bounded disturbances is an interesting topic for future work.  

\begin{ack}                               
Financial support from the Poul Due Jensen Foundation for this research is gratefully acknowledged.   
\end{ack}

\bibliographystyle{plain}        
\bibliography{autosam}           



\appendix
\section{Some useful results from Probability theory}
We state some useful results for sake of completeness based on \cite{cesa2006prediction}. Consider a probability space $(\Omega, \mathcal{F}, \mathbb{P})$ where $\Omega$ is the sample space, $\mathcal{F}$ is the increasing $\sigma$-algebra and $\mathbb{P}$ is the probability measure on the measurable space $(\Omega, \mathcal{F})$. We can now define the Martingale Difference Sequences.
\begin{defn}[Martingale Difference Sequence]\label{defn:MDS}
    A sequence of random variables $X_1, X_2, \cdots$ is a Martingale Difference sequence with respect to the filtration $\{\mathcal{F}_0, \mathcal{F}_1, \cdots\}$ if for every $t>0$, $X_t$ is $\mathcal{F}_t$-measurable and
    \begin{align*}
        \mathds E_{\mathbb{P}} [X_{t+1} \mid \mathcal{F}_t] = 0, \text{ almost surely.}
    \end{align*}
\end{defn}
We now state the Azuma–Hoeffding inequality which is essentially a concentration inequality that helps in deriving probabilistic bounds for a random process. 
\begin{thm}[Azuma–Hoeffding inequality]\label{th:Azuma}
    Let $X_0, X_1, \cdots$ be a Martingale Difference Sequence with respect to the filtration $\{\mathcal{F}_0, \mathcal{F}_1, \cdots\}$ with $X_t \in [Z_t, Z_t + c_t]$ for some random variable $Z_t$ which is also measurable with respect to the filtration $\{\mathcal{F}_0, \cdots, \mathcal{F}_{t-1}\}$ and constants $0 < c_1, c_2, \cdots c_t, \cdots < \infty$. Let $S_t = \sum_{t=1}^T X_t$, than for any $m>0$ we have the following inequality
    \begin{align*}
        \mathbb{P}[S_t > m] \leq \text{exp} \left( \frac{-2m^2}{\sum_{t=1}^T c_t^2} \right)
    \end{align*}
\end{thm}
We now state the Borel–Cantelli Lemma which reasons about the probability of events occurring infinitely often.  
\begin{lem}[Borel–Cantelli Lemma]\label{lemma:Borel-Cantelli}
    Let $E_1,E_2, \cdots$ be a sequence of events in some probability space. If 
    \begin{align*}
        \sum\limits_{t = 1}^\infty \mathbb{P}[E_t] < \infty,
    \end{align*}
    then the probability that infinitely many of them occur is
    \begin{align*}
        \mathbb{P}[\limsup\limits_{t \to \infty} E_t] = 0.
    \end{align*}
\end{lem}

\end{document}